%% file: main.tex
\documentclass[a4paper,11pt]{article}
\pdfoutput=1 
\usepackage{jinstpub}
\usepackage[english]{babel}
\input{packages}

\usepackage{layouts} 
\usepackage{caption}
\usepackage{subcaption}
\DeclareSIUnit[]\gforce{\text{\ensuremath{g_{\textup{0}}}}}

\definecolor{mycolor0}{RGB}{ 31, 119, 180}  
\definecolor{mycolor1}{RGB}{255, 127,  14}  
\definecolor{mycolor2}{RGB}{ 44, 160,  44}  
\definecolor{mycolor2b}{RGB}{116,196,118}
\definecolor{mygraydark}{gray}{0.1}  
\definecolor{mygray}{gray}{0.4}  
\definecolor{mygraylight}{gray}{0.6}  
\colorlet{arrowcolor}{mycolor0}

\title{\boldmath STRAW-b (STRings for Absorption length in Water-b): the second pathfinder mission for the Pacific Ocean Neutrino Experiment}

\author[a, 1]{K. Holzapfel, \note{Corresponding author.}}
\author[a]{C. Spannfellner,}
\author[b]{O. Aghaei,} 
\author[b]{A. Baron,} 
\author[b]{J. Bedard,} 
\author[a]{M. Böhmer,}
\author[b]{J. Bosma,} 
\author[b]{N. Deis,} 
\author[a]{C. Fink,}
\author[a]{C. Fruck,}
\author[c]{A. Gärtner,}
\author[a]{R. Gernhäuser,}
\author[d]{F. Henningsen,}
\author[b]{R. Hotte,} 
\author[b]{R. Jenkyns,} 
\author[a]{M. Karl,}
\author[e]{Na. Khera,} 
\author[f]{Ni. Khera,} 
\author[b]{I. Kulin,} 
\author[b]{A. Lam,} 
\author[b]{T. Lavallee,} 
\author[a]{K. Leismüller,}
\author[a]{L. Papp,}
\author[b]{B. Pirenne,} 
\author[b]{E. Price,} 
\author[b]{T. Qiu,} 
\author[g]{I. C. Rea,}
\author[a]{E. Resconi,}
\author[b]{A. Round,} 
\author[h,i]{C. Rott,} 
\author[b]{A. Ruskey,} 
\author[a]{L. Ruohan,}
\author[b]{K. Sasaki,} 
\author[b]{M. Tradewell,} 
\author[j]{M. Traxler,}
\author[g, k]{D. Vivolo,}
\author[b]{S. Wagner,} 
\author[a]{E. L. Winter,}
\author[a]{M. Wolf,}

\affiliation[a]{Department of Physics, Technical University of Munich, Germany}
\affiliation[b]{Ocean Networks Canada, University of Victoria, Victoria, British Columbia, Canada}
\affiliation[c]{Department of Physics, University of Alberta, Edmonton, Alberta, Canada} 
\affiliation[d]{Department of Physics, Simon Fraser University, Burnaby, Canada} 

\affiliation[e]{Center for Advancing Electronics Dresden, Technische Universität Dresden, Dresden, Germany} 
\affiliation[f]{Department of Physics, Technical University Kaiserslautern, Kaiserslautern, Germany} 
\affiliation[g]{National Institute of Nuclear Physics, Complesso universitario di Monte Sant'Angelo, Napoli, Italy} 
\affiliation[h]{Department of Physics and Astronomy, University of Utah, Salt Lake City, USA} 
\affiliation[i]{Department of Physics, Sungkyunkwan University, Suwon, South Korea} 
\affiliation[j]{GSI Helmholtz Centre for Heavy Ion Research, Darmstadt, Germany} 
\affiliation[k]{Department of Mathematics and Physics, University of Campania “Luigi Vanvitelli”, Caserta, Italy}

\emailAdd{kilian.holzapfel@tum.de}
\emailAdd{christian.spannfellner@tum.de}

\abstract{\input{abstract}}

\keywords{Neutrino detectors; Cherenkov detectors; Photon detectors; Detectors for astroparticle physics; Particle tracking detectors}

\arxivnumber{2310.16714} 

\date{February 6, 2024}

\colorlet{myarrowcolor0}{mycolor0!75!white}
\begin{document}

\maketitle

\input{intro}

\subimport{strawb}{strawb}

\subimport{deployment}{deployment}

\input{module_operation}
\subimport{daq}{daq}
\input{software_package}
\input{summary_outlook}

\section*{Acknowledgments}
The authors are grateful and appreciative of the support provided by Ocean Networks Canada, an initiative of the University of Victoria funded in part by the Canada Foundation for Innovation. This work is supported by the German Research Foundation through grant SFB 1258 “Neutrinos and Dark Matter in Astro- and Particle Physics” and the ORIGINS Excellence Cluster.

\input{abbreviations}

\bibliography{bib}

\end{document}

%% file: packages.tex
\usepackage{url}

\addto\extrasenglish{\renewcommand{\chapterautorefname}{Chapter}}
\addto\extrasenglish{\renewcommand{\sectionautorefname}{Section}}
\addto\extrasenglish{\renewcommand{\subsectionautorefname}{Section}}
\addto\extrasenglish{\renewcommand{\subsubsectionautorefname}{Section}}

\newcommand{\Autoref}[1]{\begingroup\def\figureautorefname{Figure}\def\figurenamautorefname{Figure}\def\tableautorefname{Table}\def\partautorefname{Part}\def\appendixautorefname{Appendix}\def\equationautorefname{Equation}\def\Itemautorefname{Item}\def\chapterautorefname{Chapter}\def\sectionautorefname{Section}\def\subsectionautorefname{Section}\def\subsubsectionautorefname{Section}\def\paragraphautorefname{Paragraph}\def\Hfootnoteautorefname{Footnote}\def\AMSautorefname{Equation}\def\theoremautorefname{Theorem}\def\pageautorefname{Page}\autoref{#1}\endgroup}

\usepackage[printonlyreused, nohyperlinks, nolist]{acronym}  
\newcommand{\aci}[1]{\acsu{#1} (\acl{#1})} 
\newcommand{\ach}[1]{\acs{#1}} 

\usepackage{siunitx} 
\DeclareSIUnit\pixel{pixel}
\DeclareSIUnit\permille{\text{\textperthousand}}
\DeclareSIUnit[number-unit-product = {}]{\inch}{''}
\DeclareSIUnit[number-unit-product = {}]{\foot}{\mbox{'}}

\usepackage{import} 

\usepackage{stackengine}

\usepackage{tikz}
\usetikzlibrary{arrows,shapes,positioning,  quotes, angles, math, patterns, patterns.meta, calc,trees}
\usetikzlibrary{arrows.meta,babel}

\usepackage{wrapfig}
\usepackage{graphicx}
\usepackage{floatrow}
\usepackage{booktabs}
\usepackage{xltabular}  

\usepackage[switch]{lineno} 
\usepackage{csquotes} 

\usepackage{setspace}

%% file: abstract.tex
Since 2018, the potential for a high-energy neutrino telescope, named the \ac{pone}, has been thoroughly examined by two pathfinder missions, STRAW and STRAW-b, short for \aclu{straw}\acused{strawb}. The \ac{pone} project seeks to install a neutrino detector with a one cubic kilometer volume in the Cascadia Basin's deep marine surroundings, situated near the western shores of Vancouver Island, Canada. To assess the environmental conditions and feasibility of constructing a neutrino detector of that scale, the pathfinder missions, STRAW and STRAW-b, have been deployed at a depth of \SI{2.7}{\kilo\meter} within the designated site for P-ONE and were connected to the NEPTUNE observatory, operated by \ac{onc}. While \ac{straw} focused on analyzing the optical properties of water in the Cascadia Basin, \ac{strawb} employed cameras and spectrometers to investigate the characteristics of bioluminescence in the deep-sea environment.
This report introduces the \ac{strawb} concept, covering its scientific objectives and the instrumentation used. Furthermore, it discusses the design considerations implemented to guarantee a secure and dependable deployment process of \ac{strawb}. Additionally, it showcases the data collected by battery-powered loggers, which monitored the mechanical stress on the equipment throughout the deployment. The report also offers an overview of \ac{strawb}'s operation, with a specific emphasis on the notable advancements achieved in the \ac{daq} system and its successful integration with the server infrastructure of \ac{onc}. 


%% file: intro.tex
\section{Introduction}
\begin{figure}
  \begin{subfigure}[c]{\textwidth}
    \includegraphics[width=\textwidth]{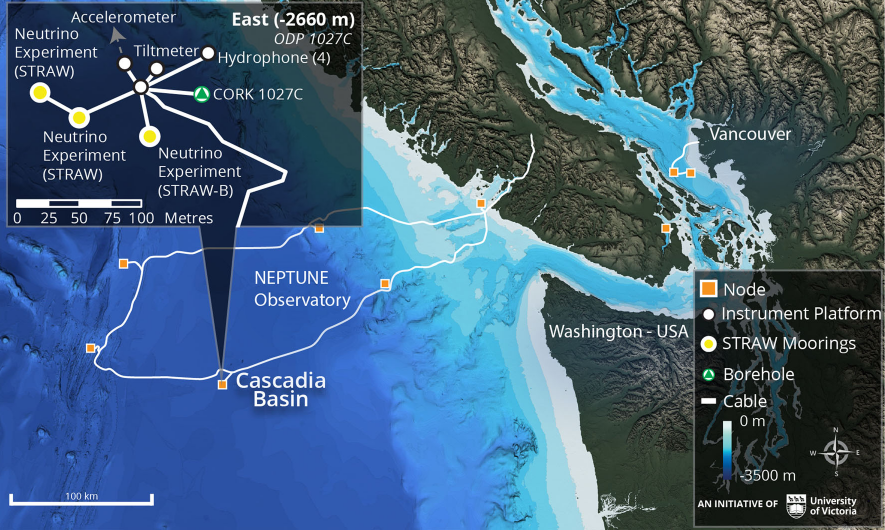}
  \end{subfigure}
\caption{Maps of ONC's NEPTUNE Observatory. P-ONE will be located at the Cascadia Basin node, where the pathfinder projects STRAW and STRAW-b have been until their recovery in July 2023. Image by courtesy of \ac{onc}.}
\label{intro:fig:neptune}
\end{figure}

Astrophysical neutrino observations require additional neutrino telescopes on a cubic-kilometer scale alongside the operational IceCube detector \cite{ngc1068, galacticplane2023, txs2018}. Notable examples for such telescopes include \aci{km3net} \cite{km3net2012}, \aci{gvd} \cite{gvd2019}, and \aci{pone} \cite{pone2023}. 
\ac{pone} is set to be situated within the Cascadia Basin, located off the shore of Vancouver Island, Canada, as shown in \autoref{intro:fig:neptune}. This initiative will leverage the existing deep-sea infrastructure, NEPTUNE, managed by \ac{onc} \cite{onc2013}. The project aims to instrument the ocean, spanning a \SI{1}{\kilo\meter} height volume from the seafloor at \SI{2660}{\meter}. Due to the challenging environmental conditions and the remote nature of the deep-sea location, a sturdy design for all components is imperative, along with comprehensive testing procedures.

Consequently, two pathfinder missions, namely \acsu{straw} and \acsu{strawb},
short for Strings for Absorption Length in Water, have been deployed at the Cascadia Basin in 2018 and 2020 to investigate environmental parameters and feasibility for the cubic-kilometer scale neutrino detector \cite{straw2019, Bailly2021}. These missions are focused on investigating environmental variables and the viability of deploying a cubic-kilometer scale neutrino detector \cite{straw2019, Bailly2021}. While \ac{straw} concentrated on assessing the optical characteristics of water in the Cascadia Basin \cite{Bailly2021}, its successor, \ac{strawb}, employed cameras and spectrometers to examine the properties of bioluminescence \cite{harvey1957, Latz1988, Gouveneaux2016, Herring1983} in the deep-sea environment.

%% file: strawb/strawb.tex
\section{STRAW-b Concept and Instrumentation}
\label{strawb:sec:strawb}

\begin{figure}
\begin{subfigure}[c]{.33\textwidth}
    \input{strawb-drawing}
\end{subfigure}\hfill
\begin{minipage}[c]{.5\linewidth}
  \begin{subfigure}[c]{\textwidth}
      \includegraphics[height=.46\textheight]{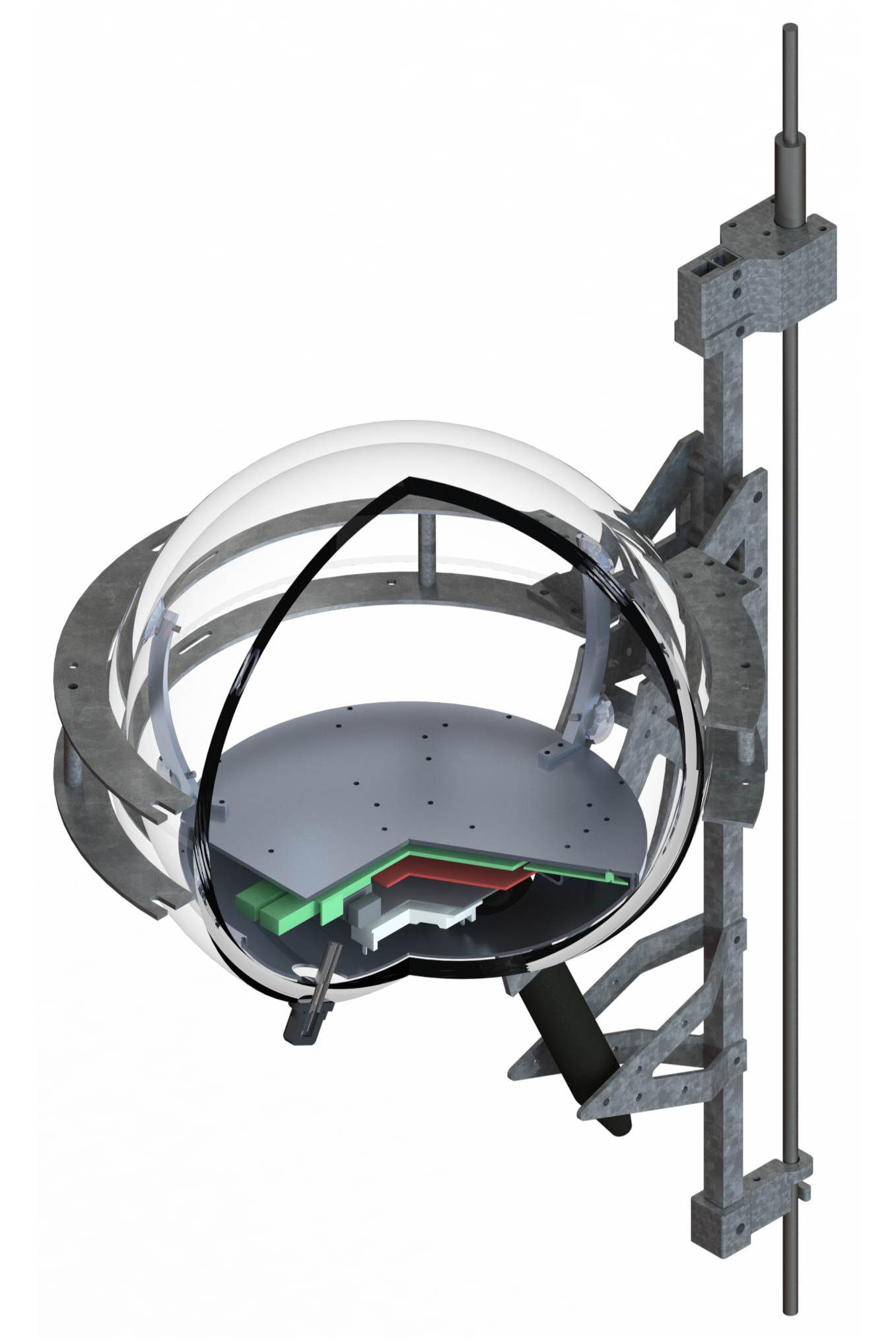}
  \end{subfigure}
  
  \begin{subfigure}[c]{\textwidth}
      \hspace{-.25cm}\includegraphics[width=.9125\textwidth]{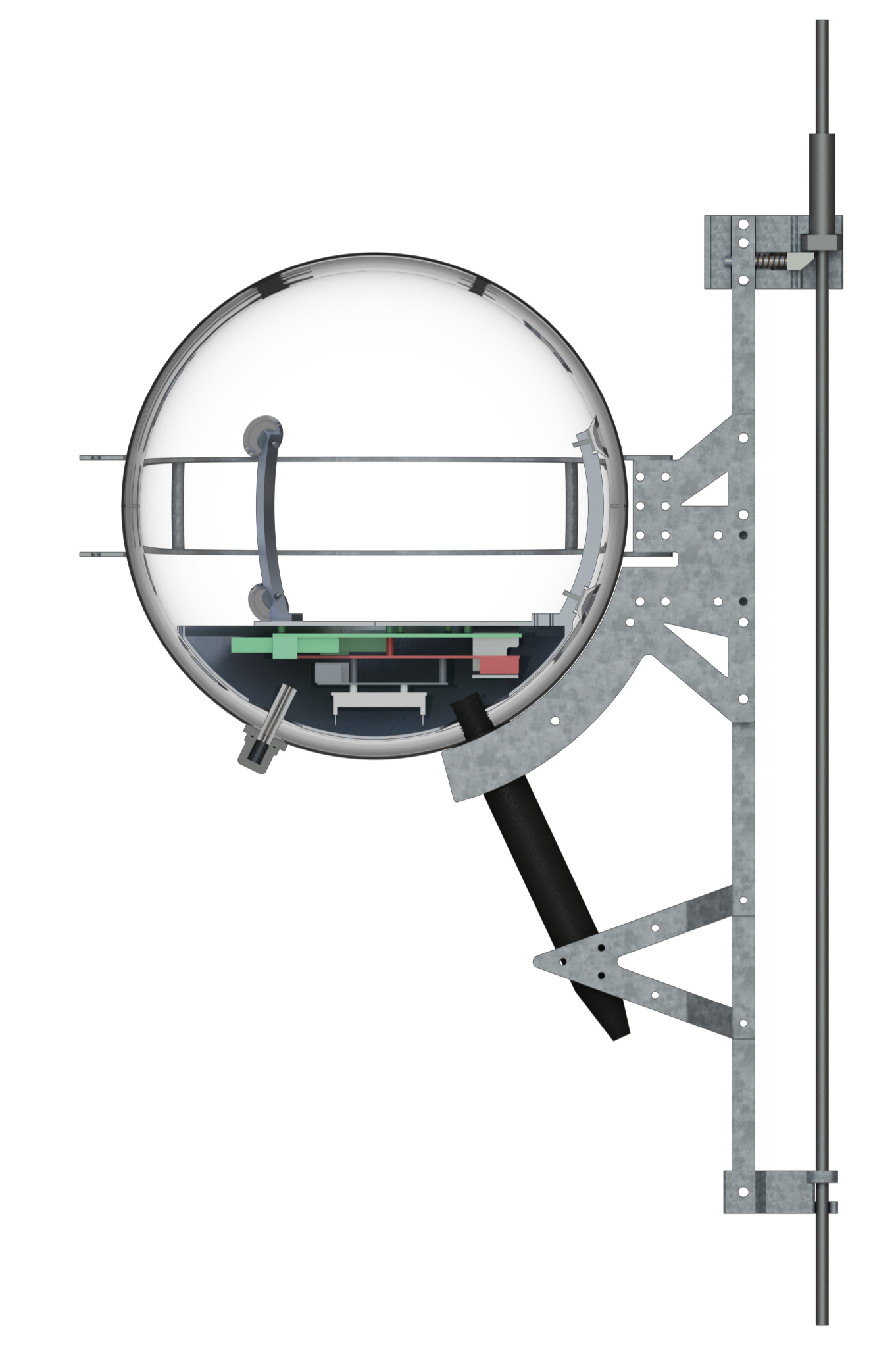}
  \end{subfigure}
  \end{minipage}
\caption[Sketch of the STRAW-b mooring]{Illustration of the STRAW-b mooring line (left), and two cut views of the standardized module (upper and lower right) where the electronics compartment is visible.
}
\label{strawb:fig:strawb}
\end{figure}

\ac{strawb}, the second pathfinder for \ac{pone}, was deployed in 2020 and situated approximately \SI{40}{\meter} East of \ac{straw} (see \autoref{intro:fig:neptune}). The primary motivation behind \ac{strawb} was to enhance our understanding of the expected background, explicitly focusing on bioluminescence. Additionally, it aimed to accumulate valuable experience for constructing and successfully deploying longer mooring lines required for extensive neutrino detector arrays.

The \ac{strawb} setup comprises ten modules mounted on a 444-meter-long mooring line, as illustrated in \autoref{strawb:fig:strawb}. With the exception of one module, all are enclosed in a 13-inch high-pressure-resistant glass housing. Each module is connected to its individual data cable providing network connectivity through glass fibers and power via copper wires. All cables are terminated at the \ac{mjb} located at the mooring's base \cite{spannfellner2020}.


\subsection{Standard Modules}
\label{strawb:sec:stdmodule}

\begin{figure}
\begin{subfigure}[c]{.45\textwidth}
    \input{figure_internal_electronics}
\end{subfigure}\hfill
\begin{subfigure}[c]{.51\textwidth}
    \includegraphics[width=\textwidth]{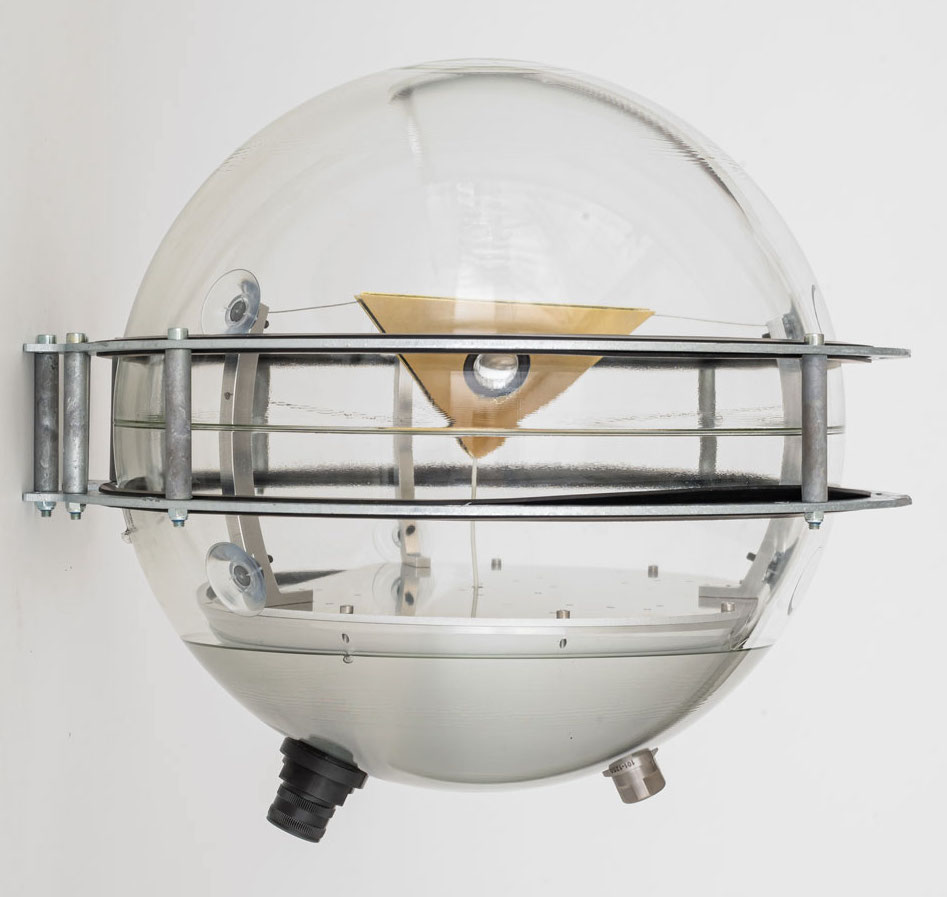}
\end{subfigure}
\caption{Standardized module electronic (left) and Standard Module 3 hosting the art \& science installation \textit{radioamnion} (right). The electronic components are positioned within the aluminum compartment located at the lower section of the pressure housing, while the upper part of the sphere is designated for the sensors of the specialized modules. The standardized module configuration is detailed in \autoref{strawb:sec:stdmodule}.}
\label{strawb:fig:standard_module}
\end{figure}

The instrumentation of \ac{strawb} followed a standardized module configuration, as shown in \autoref{strawb:fig:standard_module}. The setup integrates electronics for power management, networking, and an Odroid C2 single-board computer \cite{odroid2016}. The electronic elements are placed in the bottom section of the pressure housing, which consists of an aluminum plate and bowl. The compartment is illustrated in \autoref{strawb:fig:strawb}. To streamline the assembly process, the electronic components are securely attached to the aluminum plate, as depicted in \autoref{strawb:fig:standard_module}. The aluminum plate is then connected to the aluminum bowl. The bowl, in turn, is affixed to the bottom of the glass pressure housing using optical gel as an adhesive. This configuration also functions as a heat sink for the electronic components \cite{spannfellner2020}.

Environmental sensors were added to monitor internal pressure, temperature, and humidity, serving as checkpoints for successful deployment and ongoing monitoring of the module's condition. To track the module's orientation and movements with the currents, an accelerometer and an electronic compass were employed. To manage and monitor individual electronic subcomponents, a total of six \ac{dc} supplies are utilized, each of which is configurable and capable of measuring voltage and current. Furthermore, each module is equipped with a battery-powered data logger designed to assess the environmental conditions during the deployment. The data recorded by these loggers is presented and analyzed in \autoref{deployment:sec:logger} \cite{spannfellner2020}.

Expanding on this standardized configuration, certain modules incorporate specialized instruments tailored for specific measurements. These additional instruments are accommodated in the available space outside the electronics compartment, allowing for a modular and seamless integration. The following sections will coffer an overview of these supplementary instruments.

\begin{figure}
 \includegraphics[width=.9\textwidth]{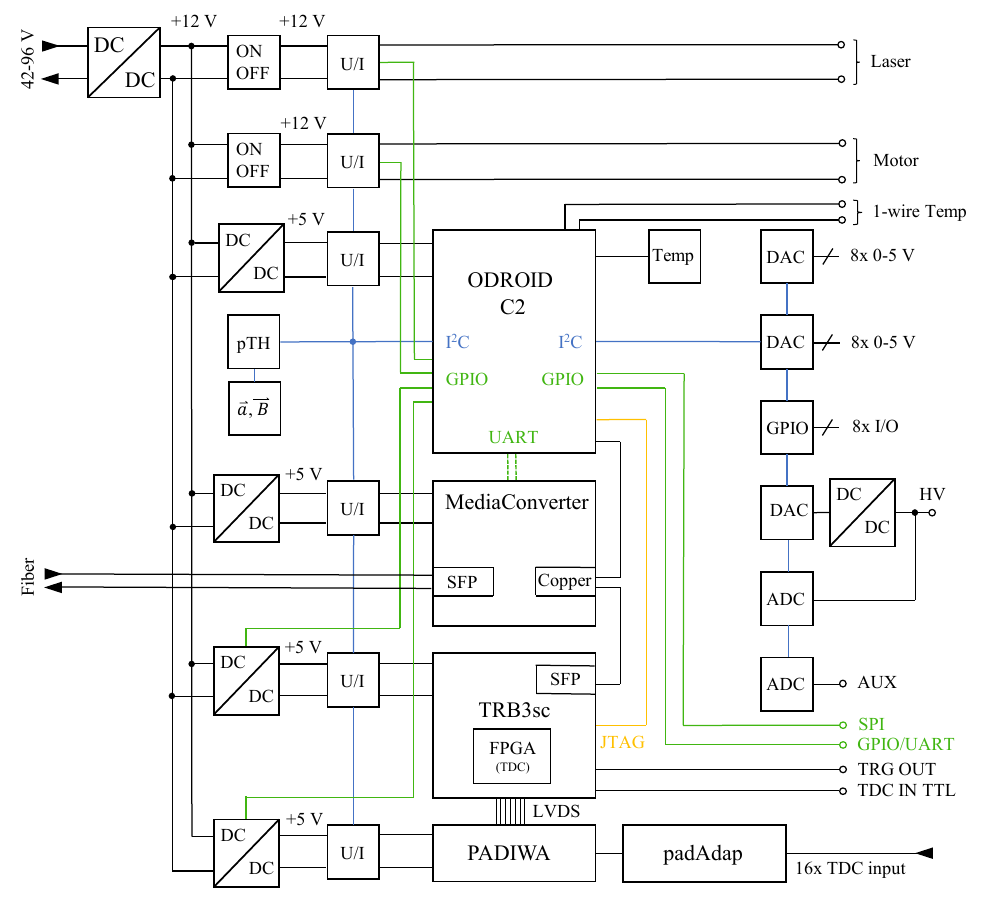}
\caption{Simplified block diagram of the electronics for the standard module. Specific functions can be tailored to the respective modules by software modifications. For more information on the base electronics refer to the text.}
\label{strawb:fig:standard_module_electronics}
\end{figure}

To ensure a reliable system, a sophisticated electronic layout has been developed, particularly as it must be adaptable to the various different configurations and dimensions. The layout is shown as a simplified block diagram in \autoref{strawb:fig:standard_module_electronics}. Data transfer and power connection between the modules and the \ac{mjb} is provided by single hybrid cables with copper wires for power and two single mode fibres for communication. At the heart of the readout is the TRB3sc (see also \autoref{daq:sec:fast_signal}), an integrated readout system based on a \ac{fpga} with central trigger system, slow control, Gigabit Ethernet and time-to-digital converter (TDC) \cite{trb}. This device was developed for collider experiments at GSI in Darmstadt and therefore benefits from experience and proven reliability \cite{trb}. The front-end of the TRB3sc is the so-called Padiwa, which amplifies and shapes the signal of the 16 inputs to allow a proper time-over-threshold measurement. The photosensor signal of each specialised module is fed to the Padiwa via the padAdap, a small adapter board. Communication with the devices is handled by an Odroid C2 \cite{odroid2016}. The connection between the C2 and the TRB3sc is made by a carrier board called Phobos. The Phobos also contains power conversion via switchable DC-DC converters, as well as several digital-to-analogue converters (DACs) and analogue-to-digital converters (ADCs). It also hosts environmental sensors for pressure, temperature, humidity, acceleration and magnetic field, which are read by the Odroid via an I2C connection. Additionally, the power consumption of each DC-DC converter can be monitored via I2C. As the connection to the Odroid is based on Ethernet, a media converter is used as the interface to the fibre optic connection. A small add-on board replaces the microcontroller on the media converter, turning it into a configurable Ethernet switch. Port-based VLAN is implemented as a default configuration, isolating the TRB3sc Ethernet port from the small form-factor pluggable (SFP) uplink. This avoids congestion of the data links by hardware or software failures and ensures the accessibility of the module. \autoref{strawb:fig:standard_module_electronics} shows the general readout structure with the subsystems mentioned. In addition, a battery-powered data logger (Uboot) is housed in the modules to specifically record the acceleration and orientation of the instruments during deployment. A magnet attached to the glass housing must be removed just prior to deployment to switch a magnetic relay and initiate data acquisition by the Uboot. The data collected by the Uboot during the deployment is presented in \autoref{sec:deployment} \cite{spannfellner2020}.

\subsection{LiDAR Modules}
\label{strawb:sec:lidar}

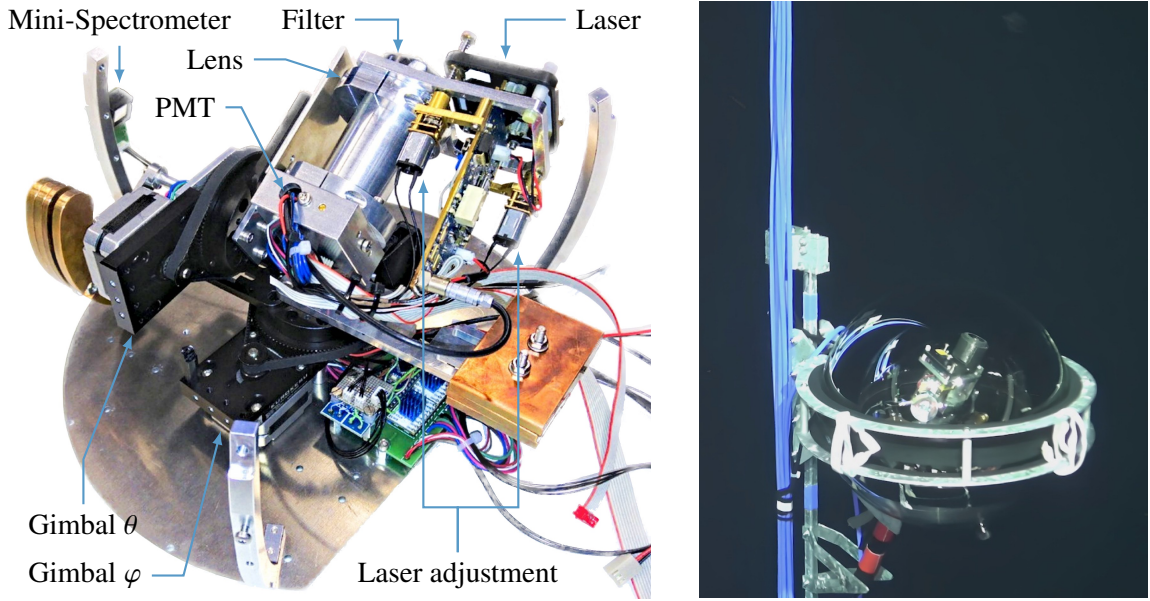
\begin{figure}
\begin{subfigure}[c]{.55\textwidth}
  \begin{tikzpicture}[every label/.append style={text=black, font=\normalsize},
                                every node/.append style={font=\normalsize}]
    \node[anchor=south west,inner sep=0] (image) at (0,0) {\includegraphics[width=\textwidth]{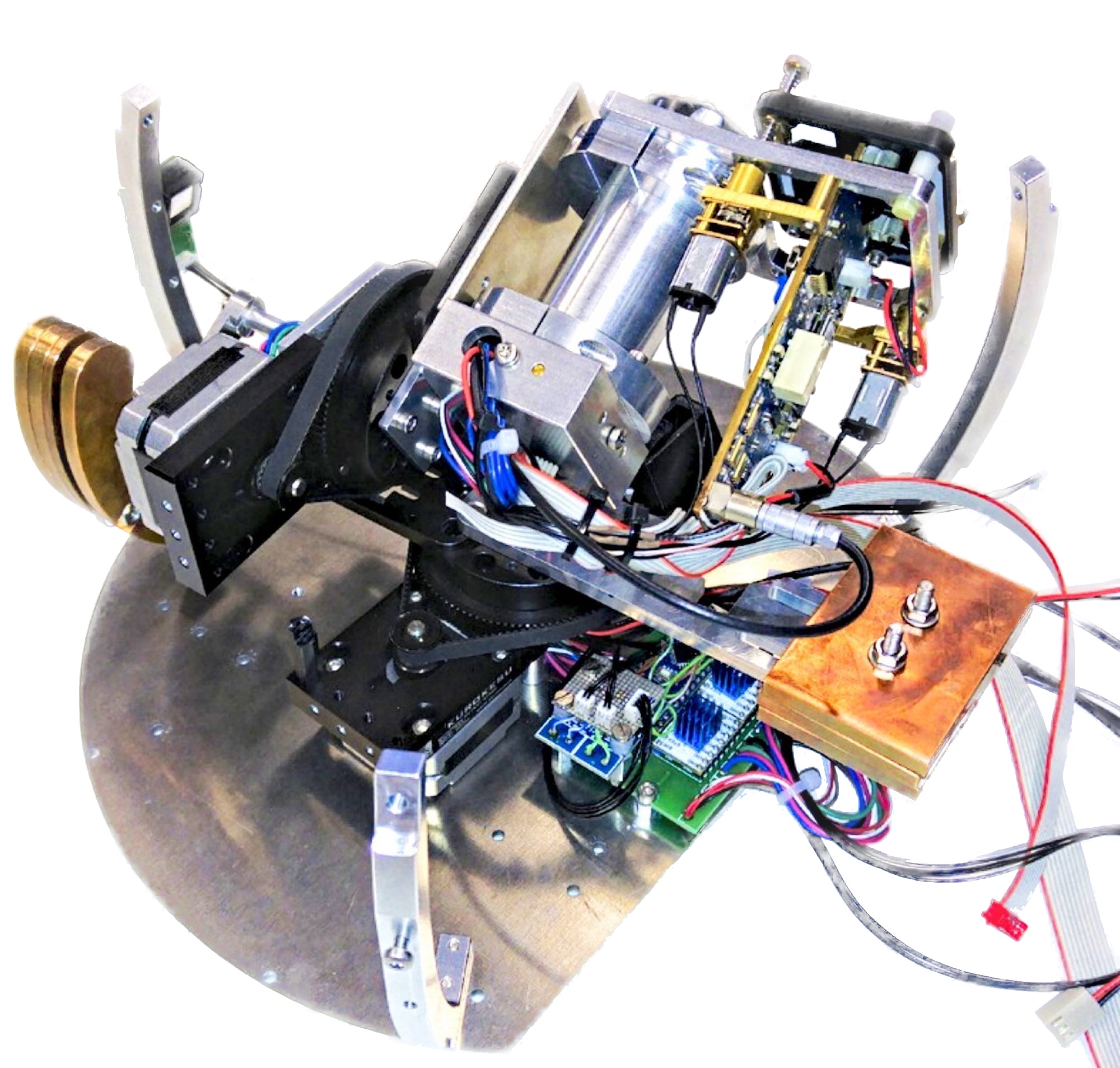}};
    \begin{scope}[x={(image.south east)}, y={(image.north west)}]
    	\node [anchor=north east] (laser) at (1,1) {Laser};
    	\draw [-latex, thick, myarrowcolor0] (laser.west) -| (.77,.9);
    	
    	\node [anchor=north west] (filter) at (.4,1) {Filter};
    	\draw [-latex, thick, myarrowcolor0] (filter.east) -| (.6,.91);
    	
    	\node [anchor=west] (lens) at (.25,.9) {Lens};
    	\draw [-latex, thick, myarrowcolor0] (lens.east) -- +(.05,0) -\ (.51,.87);
    	
    	\node [anchor=north] (minispec) at (.16,1) {Mini-Spectrometer};
    	\draw [-latex, thick, myarrowcolor0] (minispec.south) -| (.16,.86);
    	
    	\node [anchor=west] (pmt) at (.2,.82) {PMT};
    	\draw [-latex, thick, myarrowcolor0] (pmt.east) -- +(.025,0) -\ (.42,.68);
    	
    	\node [anchor=south west] (lasermotor) at (.52,0) {Laser adjustment};
    	\draw [-latex, thick, myarrowcolor0] (lasermotor.north) -- +(0,.075) -| (.64,.7);
    	\draw [-latex, thick, myarrowcolor0] (lasermotor.north) -- +(0,.075) -| (.79,.58);
    	
    	\node [anchor=south west] (gimbal2) at (0,.09) {Gimbal $\theta$};
    	\draw [-latex, thick, myarrowcolor0] (gimbal2.north) -- +(0,.075) -\ (.18,.44);
    	
    	\node [anchor=south west] (gimbal) at (0,0) {Gimbal $\varphi$};
    	\draw [-latex, thick, myarrowcolor0] (gimbal.east) -- +(.05,0) -\ (.32,.28);
    \end{scope}
  \end{tikzpicture}
\end{subfigure}\hfill
\begin{subfigure}[c]{.39\textwidth}
    \includegraphics[width=\textwidth]{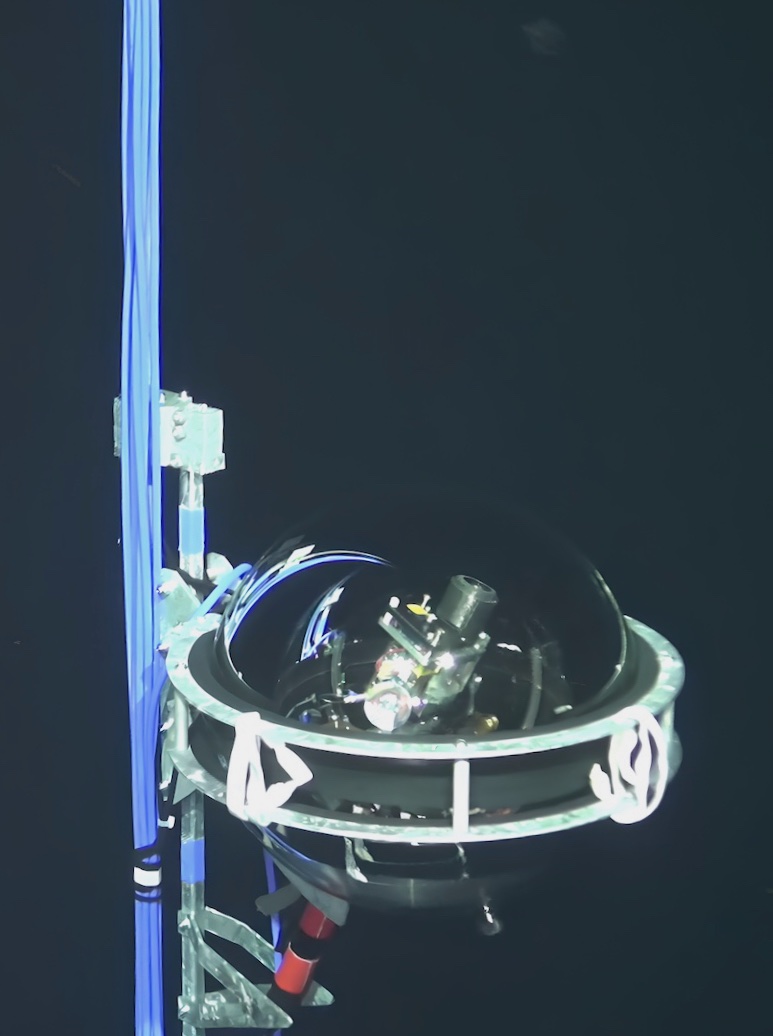}
\end{subfigure}
\caption[Picture of the LiDAR]{Picture of the internal LiDAR components (left) and during an inspection dive with a \ac{rov} (right). In the image of the inspection (right), the LiDAR is pointing towards the upper right. Further information about the module is elaborated upon in \autoref{strawb:sec:lidar}.}
\label{strawb:fig:lidar}
\end{figure}

Two \ac{lidar} modules aim to verify the attenuation length measurements from \ac{straw} and separately determine scattering and absorption length. These are equipped with a pulsed laser diode as an emitter, while a lens focuses the back-scattered photons onto a micro \ac{pmt} device to measure intensity over time. A complete assembly of the internal \ac{lidar} components is depicted in \autoref{strawb:fig:lidar}. The procedural steps of the measurement are outlined below.

An electric trigger signal initiates the emission of a \SI{450(10)}{\nano\meter} light pulse by a laser diode (NPL45B) \cite{laser2019}. The light pulse is emitted after a delay of approximately \SI{35}{\nano\second} \cite{laser2019} to the trigger, which is operated at \SI{10}{\kilo\hertz}. The laser is configured to the shortest available pulse width of \SI{5(1)}{\nano\meter} \aci{fwhm} \cite{laser2019}. The back-scattered photons are then detected by a micro PMT H12406 from Hamamatsu \cite{micropmt2022}. To limit the PMT to mainly detect the laser's light, a filter \SI{450(10)}{\nano\meter} and a 1" lens is used. The lens also enhances the light sensitivity to the laser beam. Furthermore, the laser's intensity is set to a level where the PMT detects with approximately \SI{50}{\percent} at least one photon per pulse. The micro PMT H12406 has an internal counting electronic that provides an electric signal of \SI{20}{\nano\second} for each detected photon with a delay of \SI{6}{\nano\second} including the \ac{tts} \cite{micropmt2022}. The laser trigger and PMT's electric signals are digitalized by a \ac{trb} \cite{trb}, as outlined in \autoref{daq:sec:fast_signal}. The \ac{trb} has a sub-nanosecond timing precision and provides two timestamps for each detected photon: the time of the laser trigger and the time delay from the laser trigger to the PMT's signal. These two values are critical for \ac{lidar} measurements, enabling the analysis to determine the number of detected photons per laser pulse and the time of flight of each detected photon \cite{fink2019, spannfellner2020}.

The laser, filter, lens, and \ac{pmt} are affixed to a rotation stage featuring two axes. This configuration permits adjustments to the orientation in azimuth and zenith. Furthermore, two motors enable alterations to the laser's pointing direction, ensuring precise alignment with respect to the \ac{fov} of the PMT \cite{fink2019, spannfellner2020}.

The LiDAR is equipped with an in-situ laser calibration system. To achieve this, the gimbal is positioned in a specific manner, allowing the laser to illuminate both a pin-diode and a C12880MA mini-spectrometer from Hamamatsu \cite{minispec2023} simultaneously. This mini-spectrometer corresponds to the type employed in the Mini-Spectrometer and PMT-Spectrometer modules. Through this configuration, it becomes possible to measure both the spectrum and the intensity of the laser during its lifetime in the deepsea \cite{fink2019, spannfellner2020}.

\subsection{Muon-Tracker Module}
\label{strawb:sec:muontracker}

\begin{figure}
\begin{subfigure}[c]{.55\textwidth}
 \begin{tikzpicture}[every label/.append style={text=black, font=\normalsize},
                                every node/.append style={font=\normalsize}]
    \node[anchor=south west,inner sep=0] (image) at (0,0) {\includegraphics[width=\textwidth, trim={0 0 1.2cm 0},clip]{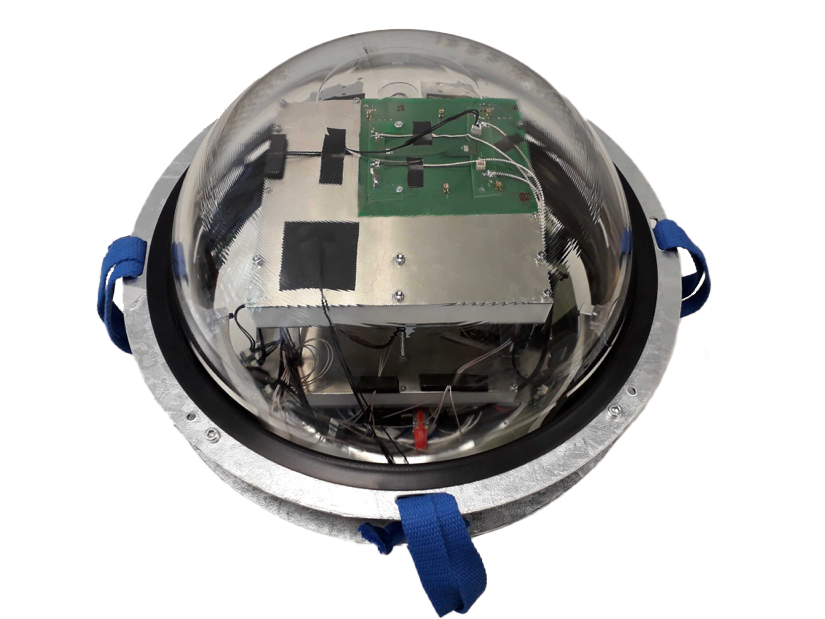}};
    \begin{scope}[x={(image.south east)}, y={(image.north west)}]
        \def\xl{0}
        \draw [-, dashed, thick, mycolor2!50!white]  (.54,.5) -/ (.54,.84);
        \draw [-, dashed, thick, mycolor2!50!white]  (.36,.695) -/ (.72,.695);
        
        \node [anchor=west, align=left] (minispec) at (\xl,.78) {Upper\\Scintillators};
        \draw [-{*[fill=mycolor2!50!white]}, thick, mycolor2] (minispec.east) -| (.45,.6);
        \draw [-{*[fill=mycolor2!50!white]}, thick, mycolor2] (minispec.east) -| (.64,.6);
        \draw [-{*[fill=mycolor2!50!white]}, thick, mycolor2] (minispec.east) -| (.64,.78);
        \draw [-{*[fill=mycolor2!50!white]}, thick, mycolor2] (minispec.east) -| (.45,.78);
        
        \node [anchor=north west, align=left] (sipm1) at (0,1) {SiPM Array\\Positions};
        \draw [-{*[fill=myarrowcolor0]}, thick, myarrowcolor0] (sipm1.east) -| (.52,.67);
        \draw [-{*[fill=myarrowcolor0]}, thick, myarrowcolor0] (sipm1.east) -| (.52,.81);
        \draw [-{*[fill=myarrowcolor0]}, thick, myarrowcolor0] (sipm1.east) -| (.71,.67);
        \draw [-{*[fill=myarrowcolor0]}, thick, myarrowcolor0] (sipm1.east) -| (.685,.81);
        
        \node [anchor=south east, align=left] (sipm2) at (1,0) {SiPM Array\\Positions};
        \draw [-latex, thick, myarrowcolor0] (sipm2.west) -| (.37,.465);
        \draw [-latex, thick, myarrowcolor0] (sipm2.west) -| (.55,.475);
        
        \node [anchor=west, align=left] (temp) at (\xl,.175) {Temperature\\Sensor};
        \draw [-latex, thick, myarrowcolor0] (temp.east) -| (.42,.54);
        
        \node [anchor=south west] (sci2) at (0,0) {Lower Scintillators};
        \draw [-latex, thick, myarrowcolor0] (sci2.east) -| (.5,.35);
    \end{scope}
\end{tikzpicture}
\end{subfigure}
\begin{subfigure}[c]{.44\textwidth}
\includegraphics[width=\textwidth, trim={1cm 0 0 0},clip]{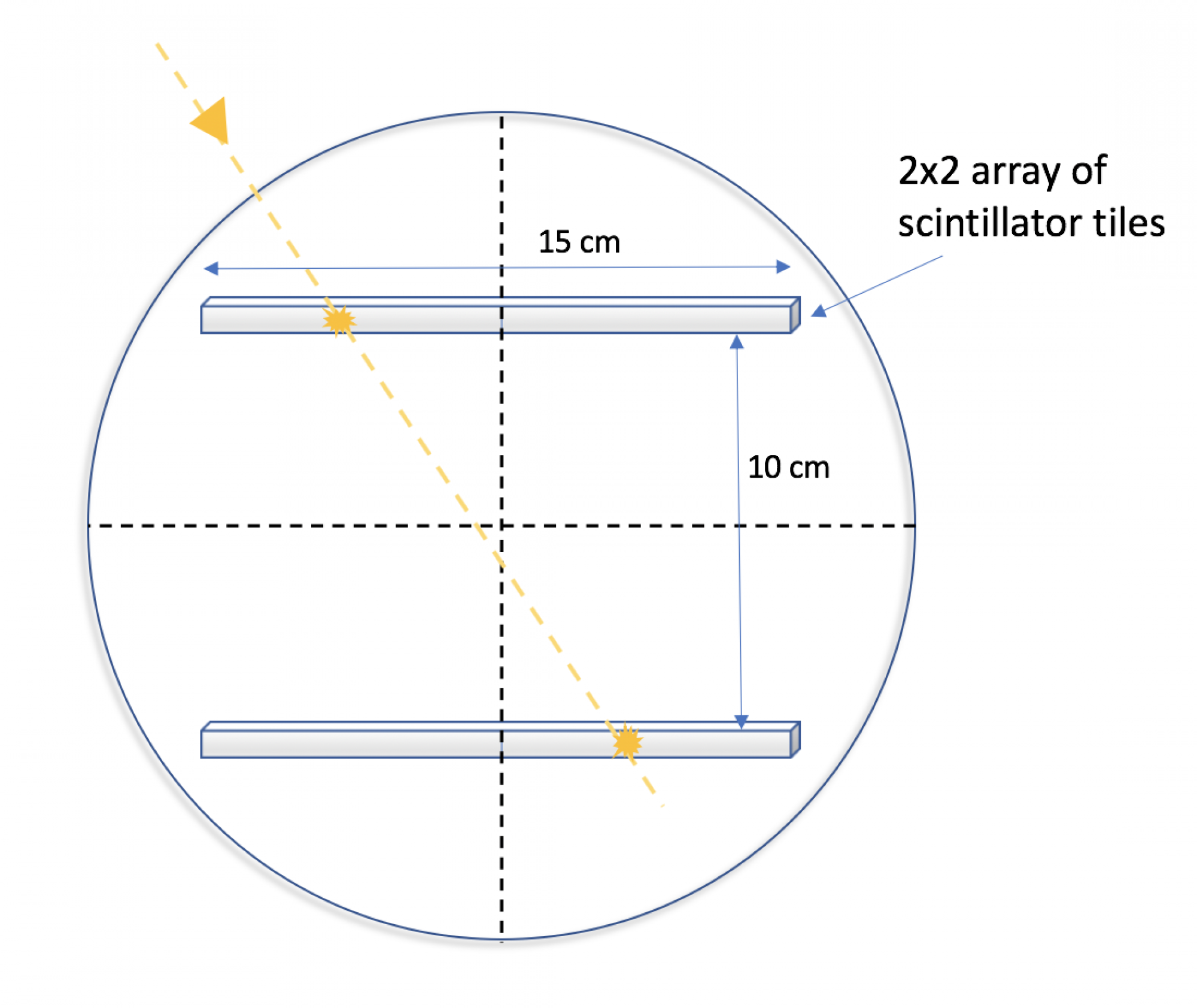}
\end{subfigure}
\caption{Picture of the Muon-Tracker module (left) and sketch with a muon track interacting in the two scintillator arrays (right). Further information about the module is elaborated upon in \autoref{strawb:sec:muontracker}.}
\label{strawb:fig:muontracker}
\end{figure}
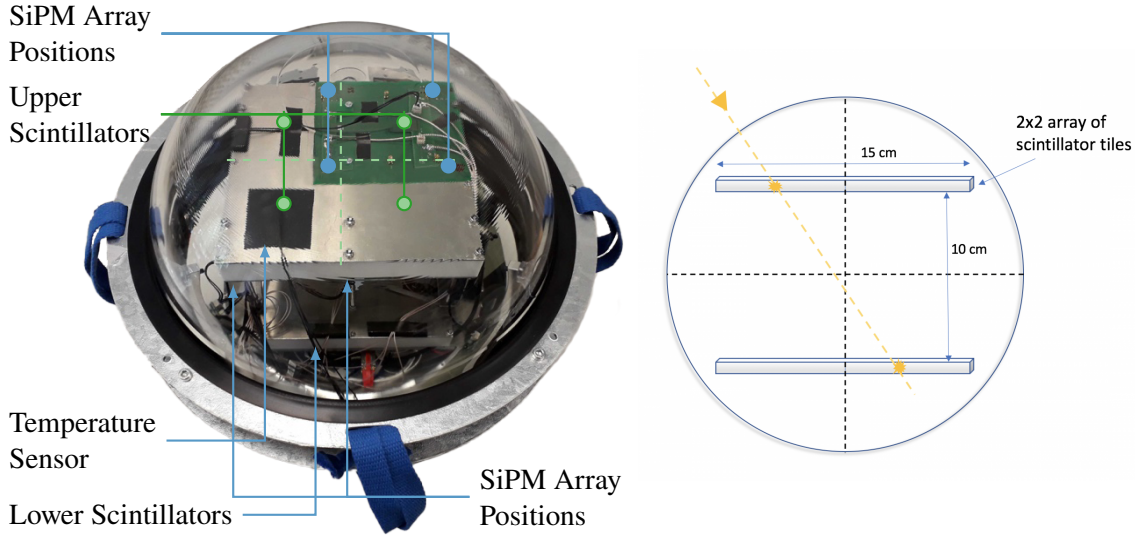

One Muon-Tracker module, depicted in \autoref{strawb:fig:muontracker}, is devised to detect muons within the confined module volume. This detection holds particular significance, as it can function as a calibration instrument, validating pointing resolution for large-scale neutrino detectors such as \ac{pone}. Upon measuring a muon, the device imposes a constraint on the muon's trajectory, necessitating it to traverse the module. This information can then be juxtaposed with the reconstructions derived from Cherenkov light data of the neutrino detector \cite{winter2019, spannfellner2020}.

The module employs eight BC-404 plastic scintillator tiles \cite{scintillator2018} organized in two 2x2 arrays, positioned vertically at a separation of \SI{10}{\centi\meter}, as illustrated in \autoref{strawb:fig:muontracker}. Each individual scintillator tile measures \qtyproduct{75 x 75 x 20}{\milli\meter} in dimension. Within the 2x2 array configuration, the scintillator tiles are divided by Teflon walls of \SI{6}{\milli\metre} thickness \cite{winter2019}.
 
When a muon traverses the scintillator, it emits light across a broad spectrum that reaches its peak at \SI{408}{\nano\meter} \cite{scintillator2018}. Two 3x3 matrices of KETEK PM3315-WB-B0 \acp{sipm} \cite{sipm2018} measure the induced light in each scintillator tile at diagonal corners. The purpose of the 3x3 matrix is to enhance the coverage of the sensitive area, and within each matrix, the nine \acp{sipm} are linked in parallel to a single readout channel. In instances of time-correlated events occurring within separate scintillator tiles, an approximation of directional information can be derived. The 16 channels of SiPM signals are digitized using a \ac{trb} \cite{trb} and Padiwa \cite{padiwa2015}, a methodology akin to other modules. The system of \ac{trb} and Padiwa is described in \autoref{daq:sec:fast_signal} \cite{winter2019, spannfellner2020, trb, padiwa2015}.

\subsection{PMT-Spectrometer Modules}
\label{strawb:sec:pmtspec}

\begin{figure}
\begin{tikzpicture}[every label/.append style={text=black, font=\normalsize},
                                every node/.append style={font=\normalsize}]
  \node[anchor=south west,inner sep=0] (image) at (0,0) {\includegraphics[width=\textwidth, trim={0 0 0   20cm},clip]{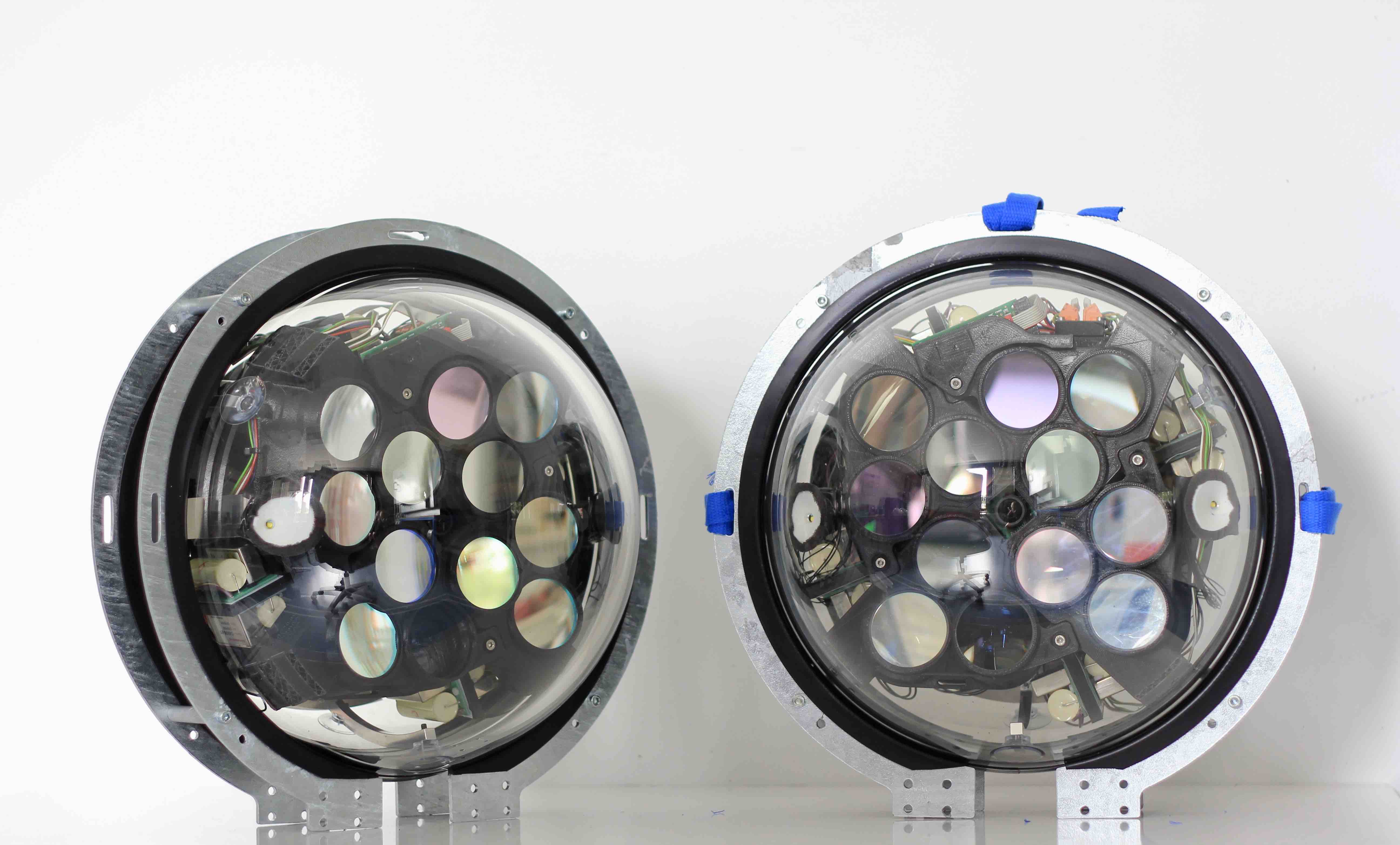}};
  \begin{scope}[x={(image.south east)}, y={(image.north west)},yscale=1.22]
    \colorlet{arrowcolor}{myarrowcolor0}
    \node [anchor=center] (camera) at (.5,.7) {Camera};
    \node[circle,thick, minimum size=5, arrowcolor] (camera_mark) at (0.72,.39) {};
    \draw [-latex, thick, arrowcolor] (camera) -| (camera_mark);
    \node[circle,thick, minimum size=5, arrowcolor] (camera_mark2) at (0.32,.38) {};
    \draw [-latex, thick, arrowcolor] (camera) -| (camera_mark2);
    
    \node [anchor=north] (minispec) at (.5,.82) {Mini-Spectrometer};
    \node[circle,thick, minimum size=5, mycolor2] (minispec_mark) at (0.685,.59) {};
    \draw [-latex, thick, arrowcolor] (minispec) -| (minispec_mark);
    \node[circle,thick, minimum size=5, mycolor2] (minispec_mark2) at (0.29,.57) {};
    \draw [-latex, thick, arrowcolor] (minispec) -| (minispec_mark2);
    
    \node [anchor=south] (nofilter) at (.5,.01) {PMTs w/o filter};
    \node[circle,thick, minimum size=5, mycolor2] (nofilter_mark) at (0.32,.24) {};
    \draw [-latex, thick, arrowcolor] (nofilter.west) -| (nofilter_mark);
    \node[circle,thick, minimum size=5, mycolor2] (nofilter_mark2) at (0.71,.25) {};
    \draw [-latex, thick, arrowcolor] (nofilter.east) -| (nofilter_mark2);
    
    \node[anchor=center] (led) at (.49,.2) {LEDs};
    \node[circle, thick,minimum size=5, mycolor2] (led_mark_l) at (0.43,0.39) {};
    \node[circle, thick,minimum size=5, mycolor2] (led_mark_r) at (0.59,0.39) {};
    \draw [-latex, thick, arrowcolor] (led) |- (led_mark_l);
    \draw [-latex, thick, arrowcolor] (led) |- (led_mark_r);
    
    \node[anchor=center] (led2) at (.05,.2) {LED};
    \node[circle, thick,minimum size=5, arrowcolor] (led_mark_2) at (0.19,0.38) {};
    \draw [-latex, thick, arrowcolor] (led2) |- (led_mark_2);
    
    \node[anchor=center] (led3) at (.95,.2) {LED};
    \node[circle, thick,minimum size=5, arrowcolor] (led_mark_3) at (0.86,0.40) {};
    \draw [-latex, thick, arrowcolor] (led3) |- (led_mark_3);
  \end{scope}
\end{tikzpicture}
\caption{The two PMT-Spectrometers post-assembly, featuring distinct sensor labeling. Circular mirrors represent the filters. Further details regarding the components and measurement principles are expounded in \autoref{strawb:sec:pmtspec}.}
\label{strawb:fig:pmtspec}
\end{figure}
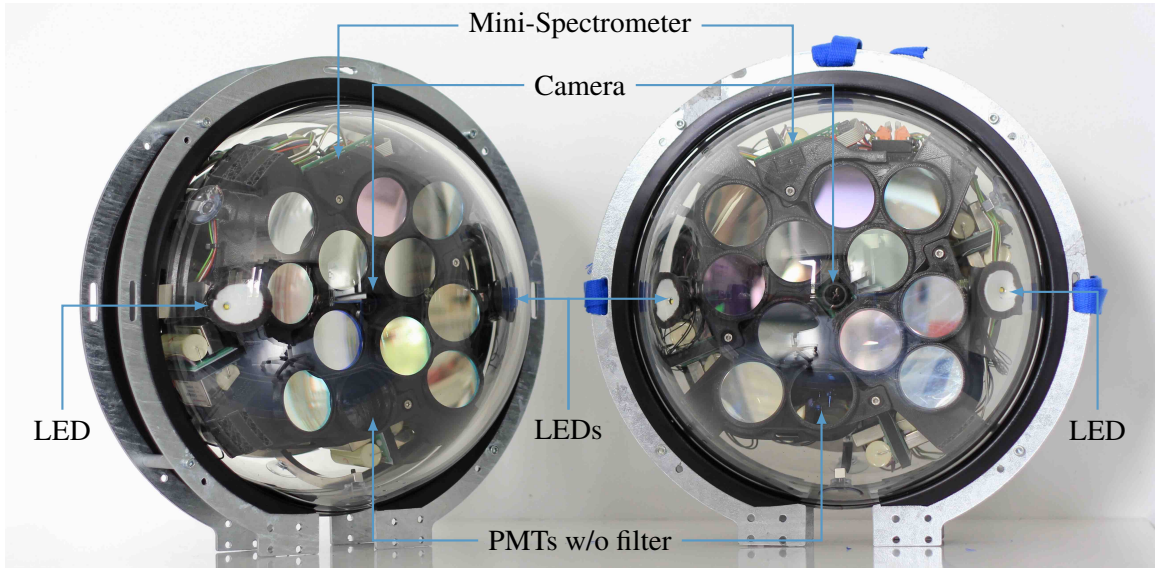

Two PMT-Spectrometer modules, illustrated in \autoref{strawb:fig:pmtspec}, are characterizing the bioluminescence emissions. Within each module, three distinct sensor types are accommodated: twelve PMTs serving as a spectrometer, a low-light color CMOS camera \cite{kang2019, arducam2023}, and a C12880MA mini-spectrometer from Hamamatsu \cite{minispec2023}. These sensors collectively enable three separate and independent bioluminescence measurements, each offering its unique benefits. All sensors are oriented upwards within the deployed modules to achieve an overlapping field of view \cite{ruohan2020, spannfellner2020, holzapfel2023}. Subsequently, a more detailed discussion of the individual sensors follows.

Each of the twelve \SI{1}{\inch} PMTs \cite{r1924pmt2014, r1925pmt1999} is paired with its own individual \SI{2}{\inch} lens \cite{lens2009} and \SI{2}{\inch} wavelength filter \cite{filter10, filter25, filter50}. The individual PMTs are deliberately separated from one another in terms of optics. The arrangement of these twelve units varies in position and angles concerning the module's vertical axis. This specific configuration counteracts the distortion caused by the spherical pressure housing of the modules while optimizing the overlapping \ac{fov} for all PMTs. The arrangement was devised through a Geant4 \cite{geant42003, geant42016} simulation, which was instrumental in determining the most advantageous positioning of these components. Given the complexity of the configuration and the constraints on available space, a 3D-printed holding structure was used to maintain the designated position of the components.
Before the light reaches a PMT, it initially passes through hard-coated OD 4.0 bandpass filters \cite{filter10, filter25, filter50}. These filters possess distinct \acp{cwl} within the range of \SIrange{350}{525}{\nano\meter} \cite{nikhera2019, ruohan2020}. It is important to note that one PMT is devoid of a filter to monitor integrated intensity. Following the filter, the light propagates through an LA1401-A lens with a focal length of \SI{59.8}{\milli\meter} $\pm$ \SI{1}{\percent} and a diameter of 50.8\scalebox{.5}{\stackon{+0.0}{-0.2}}~mm \cite{lens2009}, which effectively focuses the light onto the PMT. Two distinct types of PMTs \cite{r1924pmt2014, r1925pmt1999} are employed in order to harness their different spectral responses, where the R1924A PMT is utilized in conjunction with filters for wavelengths of \SI{510}{\nano\meter} and below, as summarized in Table \ref{strawb:tab:pmtspec}. Positioned at a distance of \SI{30}{\milli\meter} beyond the lens, the PMT is situated at approximately half the focal length. This configuration ensures the projection of a point source onto roughly half of the PMT's photocathode area. As a result, each PMT captures a distinctive spectral segment of an emission. Combined, these segments seamlessly unveil the complete spectrum of a bioluminescence emission, offering a remarkably higher sensitivity compared to the other two sensors. The signals from the 12 PMT channels are transformed into digital data through the employment of a \ac{trb} \cite{trb} and Padiwa \cite{padiwa2015} readout as described in \autoref{daq:sec:fast_signal} \cite{ruohan2020}.

The low-light camera is a custom-designed device, initially developed in partnership with Arducam for applications in the IceCube Upgrade and IceCube Gen2 \cite{kang2019, arducam2023}. This choice was motivated by the potential synergies across neutrino detectors, which necessitate hardware capable of functioning in remote locations with space constraints, low power consumption, and under low-light conditions \cite{kang2019}. Central to the camera's design is the IMX225LQR-C \acs{cmos} image sensor \cite{sony2014, sonyimx}, featuring a resolution of 1312~x~979~pixels and incorporating an RGB \ac{cfa}, often referred to as a Bayer Filter \cite{Bayer1976}. Optically, the camera utilizes an Arducam lens, specifically the M25170H12 model, with a \ac{fov} of approximately \SI{170}{\degree} \cite{arducam2023, arducamm12}. In order to further augment the camera's capabilities, an integrated LED system has been introduced into \ac{strawb}. For each camera in the PMT-Spectrometer modules, a total of six LEDs are incorporated. Within this set, two white LEDs \cite{ledwhite} are positioned in close proximity to the camera itself, as illustrated in Figure \ref{strawb:fig:pmtspec}. The remaining four LEDs are strategically located in the adjoining LiDAR modules, which are situated \SI{24}{\meter} above the PMT-Spectrometer modules, falling within the shared field of view of both the camera and the PMTs. Each LiDAR module incorporates two sets of white \cite{ledwhite} and \ac{uv} \cite{leduv} LEDs, with one set oriented downwards and the other set directed horizontally \cite{holzapfel2023}.

Furthermore, the module is also equipped with a C12880MA Mini-Spectrometer \cite{minispec2023}, akin to the configuration present in the Mini-Spectrometer module. Detailed information about this module and the C12880MA spectrometer is presented in the subsequent \autoref{strawb:sec:minispec} \cite{ruohan2020}.

\begin{table}
\caption{Summary of 12 color channels of the PMT-Spectrometer with the parameters of the filters and the type of PMT \cite{filter10, filter25, filter50, r1924pmt2014, r1925pmt1999}.}
\label{strawb:tab:pmtspec}
\begin{tabularx}{\textwidth} {rrrlXccrrrlX} \toprule
 \multicolumn{4}{c}{Lens}   &        && \multicolumn{4}{c}{Lens}   &       \\\cmidrule(lr){1-4}\cmidrule(lr){7-10}
 CWL\footnote{CWL: Center Wavelength\label{foot:cwl}} &  FWHM\footnote{FWHM: Full Width at Half Maximum\label{foot:fwhm}} & TP\footnote{TP: Transmission Peak\label{foot:tp}} &   Part & PMT &&  CWL\footref{foot:cwl} &  FWHW\footref{foot:fwhm} & TP\footref{foot:tp} &   Part & PMT \\\relax
 [nm] & [nm] & [\%] & number & type && [nm] & [nm] & [\%] & number & type \\
\cmidrule(lr){1-6}\cmidrule(lr){7-11}
   350.0 &  50.0 &  $\ge$95 & 12-099 & R1924A && 480.0 &  10.0 &  $\ge$85 & 65-206 & R1924A \\
   400.0 &  25.0 &  $\ge$90 & 86-664 & R1924A && 492.0 &  10.0 &  $\ge$85 & 65-209 & R1924A\\
   425.0 &  25.0 &  $\ge$90 & 87-799 & R1924A && 510.0 &  10.0 &  $\ge$85 & 65-213 & R1924A\\
   450.0 &  10.0 &  $\ge$85 & 65-201 & R1924A && 525.0 &  25.0 &  $\ge$90 & 87-801 & R1925 \\
   460.0 &  10.0 &  $\ge$85 & 88-016 & R1924A && 550.0 &  25.0 &  $\ge$90 & 86-667 & R1925 \\
   470.0 &  10.0 &  $\ge$85 & 65-205 & R1924A && \multicolumn{4}{c}{integrating channel without filter} & R1925\\
\bottomrule
\end{tabularx}
\end{table}

\subsection{Mini-Spectrometer Module}
\label{strawb:sec:minispec}

\begin{figure}
\begin{subfigure}[c]{.55\textwidth}
  \begin{tikzpicture}[every label/.append style={text=black, font=\normalsize},
                                  every node/.append style={font=\normalsize}]
    \node[anchor=south west,inner sep=0] (image) at (0,0) {\includegraphics[width=\textwidth, trim={0 0   1cm 0},clip]{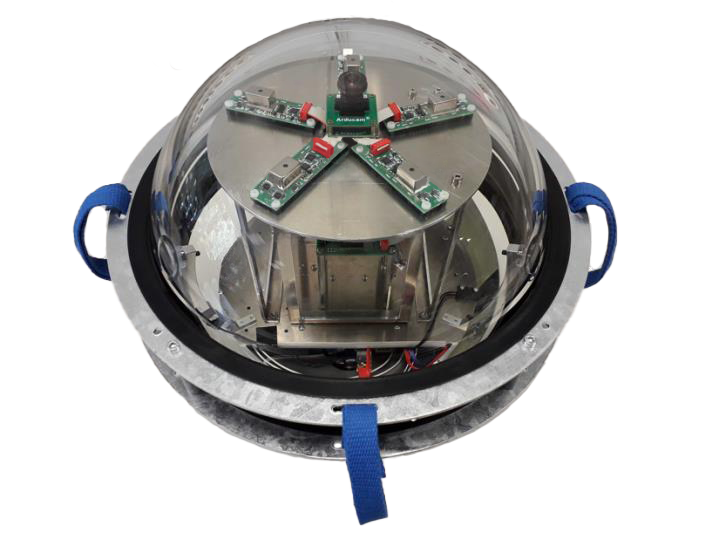}};
    \begin{scope}[x={(image.south east)}, y={(image.north west)},xscale=1.1]
  	\def\xl{0}
      \node [anchor=west] (camera) at (.75,.9) {Camera};
      \draw [-latex, thick, myarrowcolor0] (camera.west) -- (.57,.9) -/ (.53,.85);
      
      \node [anchor=west, align=left] (minispec) at (\xl,.9) {Mini-\\Spectrometer};
      \draw [-latex, thick, myarrowcolor0] (minispec.east) -- (.47,.9);
      \draw [-latex, thick, myarrowcolor0] (minispec.east) -| (.365,.84);
      \draw [-latex, thick, myarrowcolor0] (minispec.east) -- (.27, .9) |- (.39,.70);
    \end{scope}
  \end{tikzpicture}
\end{subfigure}\hfill
\begin{subfigure}[c]{.4\textwidth}
    \includegraphics[width=\textwidth]{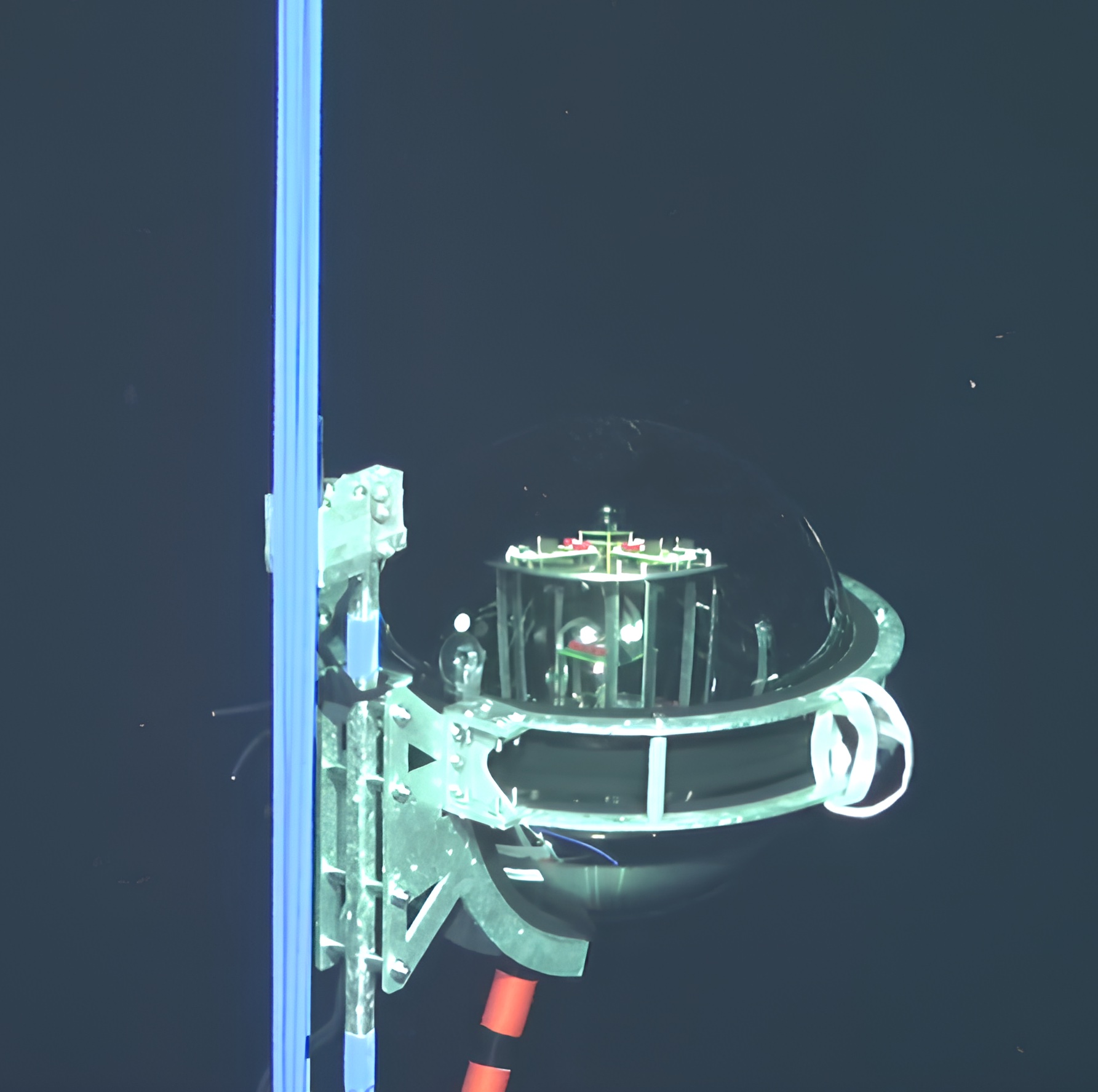}
\end{subfigure}
  \caption{Picture of the  Mini-Spectrometer Module (left) and during an inspection dive with a \ac{rov} (right). The Mini-Spectrometer Module accommodates a camera and five C12880MA Mini-Spectrometers. Further information about the module is elaborated upon in \autoref{strawb:sec:minispec}.}
  \label{strawb:fig:minispec}
\end{figure}
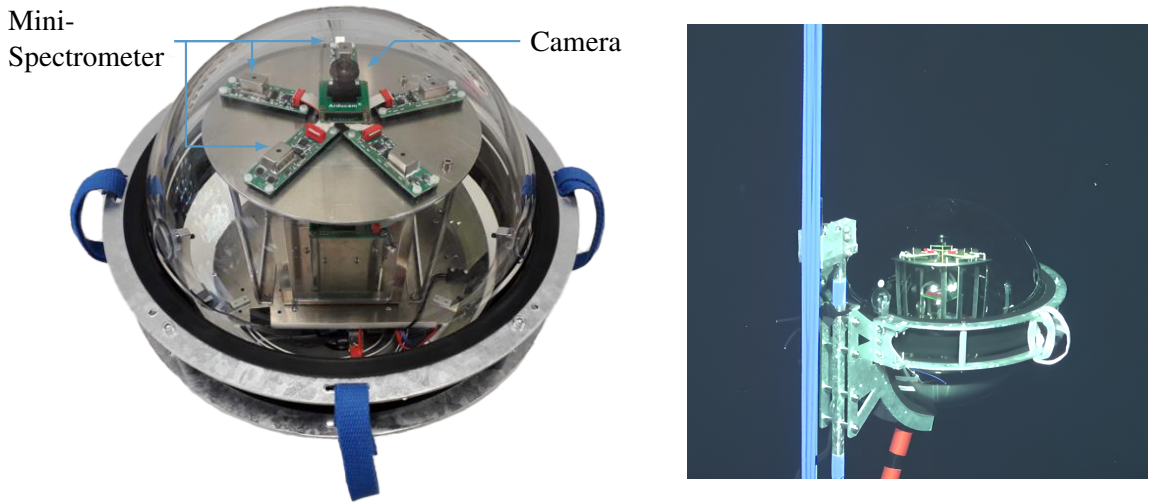

The Mini-Spectrometer module is composed of five C12880MA Mini-Spectrometers \cite{minispec2023} and a low-light camera, as depicted in \autoref{strawb:fig:minispec}. Both components are identical to the ones utilized in the PMT-Spectrometer module. The Mini-Spectrometer is a complementary device to the spectrum measurement of the PMT-Spectrometer and the cameras. It allows to measure finer resolved spectra than the PMT-Spectrometer and camera but has a lower light sensitivity. In order to offset this reduced sensitivity, the module incorporates a total of five Mini-Spectrometers. These Mini-Spectrometers are symmetrically arranged around the centrally located camera. Additionally, both the Mini-Spectrometers and the camera have their \ac{fov} oriented upwards \cite{minispec2023, rea2021, spannfellner2020}.

The measurement principle of the Mini-Spectrometer is outlined in this paragraph. The incoming light, passing through a \SI{0.5}{\milli\meter} entrance slit with a numerical aperture of 0.22, is focused by a collimating lens onto a grating chip. This grating is responsible for dispersing the incoming light into its spectral components at specific diffraction angles. The grating, combined with the chip's curvature, leads to creating an image where wavelengths are arranged linearly across the image sensor. The C12880MA Mini-Spectrometer uses a \ac{cmos} image sensor with 256 pixels to digitizes the intensities of the detected light. These pixels correspond to a spectral range spanning from \SIrange{340}{850}{\nano\meter}, providing a \ac{fwhm} resolution of less than \SI{15}{\nano\meter} \cite{minispec2023}.

\subsection{Wavelength-shifting Optical Module}
\label{strawb:sec:wom}

The \ac{wom} \cite{wom2022} is a photosensor instrument, designed to enhance the sensitivity of large detector volumes with single-photon detection capability. The key objective is to improve the signal-to-noise ratio, which is achieved by decoupling the photosensitive area from the cathode region of its \ac{pmt}. The \ac{wom} comprises a transparent tube housing two PMTs on its ends. Coated with wavelength-shifting paint that efficiently absorbs ultraviolet photons, the tube can capture and guide up to \SI{73}{\percent} (\SI{41}{\percent} in ice or water) of subsequently emitted optical photons via total internal reflection towards the PMTs. A prototype of the \ac{wom}, tailored for application in the IceCube Upgrade and IceCube Gen2 projects, was deployed as the lowermost module along the \ac{strawb} mooring line \cite{wom2022}.

\subsection{Fast Signal Digitization}
\label{daq:sec:fast_signal}

\begin{figure}
 \includegraphics[width=\textwidth]{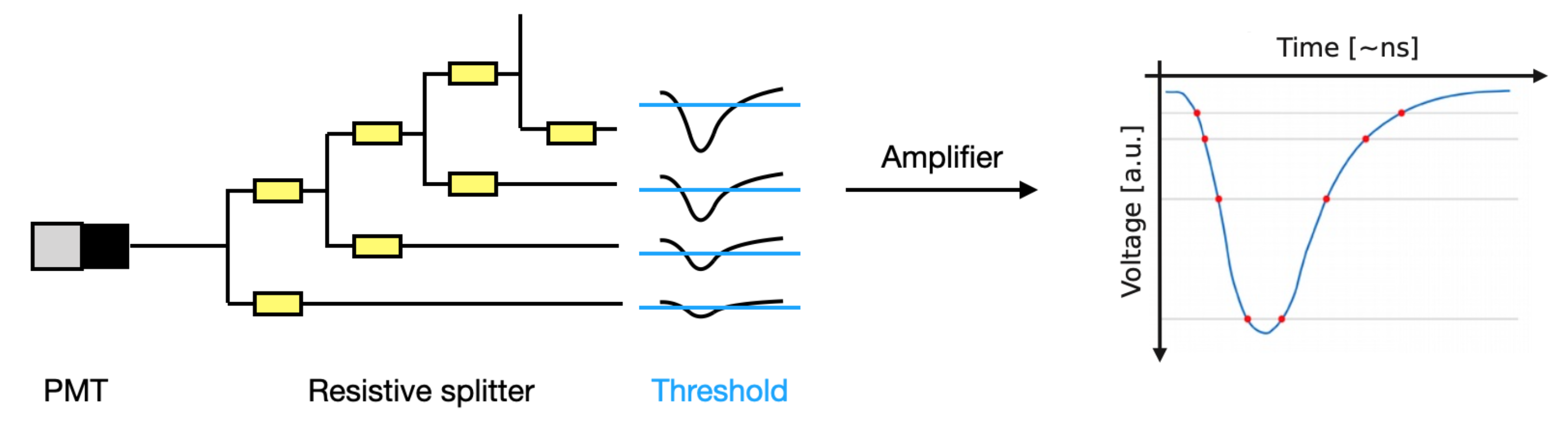}
\caption{Simplified measurement principle for the time-over-threshold measurement. The red marked positions in the left plot are the timestamps stored by the TRB with sub-nanosecond precision. For more information on the base electronics refer to the text.}
\label{strawb:fig:tot_measurement_prinziple}
\end{figure}

LiDAR, PMT-Spectrometer, and Muon-Tracker utilize a \ac{trb} (version TRB3sc) \cite{trb} to digitise signals from PMTs or SiPMs where nanosecond precision is required. The \ac{trb} can be used to perform a \ac{tot} measurement as shown in \autoref{strawb:fig:tot_measurement_prinziple}. The \ac{trb} has a 16 channel \ac{tdc} that measures the rising and falling edges of a trigger signal with sub-nanosecond precision. The \ac{tdc} is performed by firmware on a \ac{fpga}. To convert an analogue signal into a trigger signal, other electronic components are required, such as the Padiwa \cite{padiwa2015}. The Padiwa is a 16-channel discriminator board designed to enable \ac{tot} measurement with the \ac{trb}. It uses a discrete analogue stage to amplify the incoming signal, followed by a secondary slow shaping stage that allows the application of thresholds to the \ac{tot} trigger signal. The thresholds can be configured individually for each channel, with a precision of 16 bits.

An advantage of the TRB's measurement principle, with respect to full signal digitization, is effective data reduction while preserving a high time resolution \cite{trb, padiwa2015}. Furthermore, the \ac{trb} system allows the detectors to operate in two modes: high-precision and low-precision. In the high-precision mode, the exact timestamp of each pulse is recorded with sub-nanosecond precision relative to the master clock, enabling analysis at the single-photon level. In contrast, the low-precision mode records only the number of pulses within a specified time interval similar to a scaler \cite{trb}.

Storing the \ac{trb} data for long-term monitoring is not always feasible due to the large data size it generates (up to \SI{100}{\mega\byte} per minute). Therefore, the \ac{daq} operates in the low-precision mode where only the signal's threshold transgressions are counted. The counters are then recorded at a frequency of \SI{1}{\kilo\hertz}.

The signals of the SiPMs and PMTs in the Muon-Tracker and PMT-Spectrometer, respectively, are digitized with the Padiwa, where the thresholds are set at an appropriate level to achieve sensitivity to single photons.
The LiDARs are using a photon counting base for the PMT, which provides a trigger signal directly without a Padiwa. Furthermore, the LiDAR and Muon-Tracker operate in high-precision mode since both require sub-nanosecond precision, and the dark rates are optimized to a few Hz, allowing long-term monitoring in high-precision mode. In contrast, the PMT-Spectrometer operates in low-precision mode to compress the kHz to MHz signal effectively.

\subsection{Art Installations}
\label{strawb:sec:art}

In addition to its scientific objectives, specific modules of the project also serve as hosts for art installations created in collaboration with the Sonderforschungsbereich 42 (SFB42) \cite{sfb42}, a research group of artists and physicists based in Munich. This collaborative project, known as \textit{UNDERCURRENTS} \cite{undercurrents2020}, involves a group of artists from the Fine-arts Academy in Munich to facilitate a creative partnership between the two disciplines which can turn science into captivating art and forge a deeper public understanding of \ac{pone} and science, in general \cite{art2021}. As one result, the sound art project \textit{radioamnion} \cite{radioamnion2023} was developed, built, and integrated in the Standard Module 3, as depicted in \autoref{strawb:fig:standard_module}. During each full moon, \textit{radioamnion} broadcasted sonic transmissions of invited artists.

%% file: strawb/strawb-drawing.tex
\definecolor{mycolor0}{RGB}{ 31, 119, 180}  
\definecolor{mycolor1}{RGB}{255, 127,  14}  
\definecolor{mycolor2}{RGB}{ 44, 160,  44}  
\definecolor{mycolor2b}{RGB}{116,196,118}
\definecolor{mygraydark}{gray}{0.1}  
\definecolor{mygray}{gray}{0.4}  
\definecolor{mygraylight}{gray}{0.6}  

\begin{tikzpicture}[every label/.append style={text=black, font=\normalsize\linespread{0.9}\selectfont},
                                every node/.append style={font=\normalsize\linespread{0.9}\selectfont} ]
	\colorlet{mycolor0b}{mycolor0!20}
	\colorlet{mybrown}{brown!50!darkgray}
	
	\tikzmath{
		\rshift = .15; 
		\rmodule = .35;
		\rwom = \rmodule*.5;
		\zwom = \rmodule*2.5;
		\zwomshift=1./3.*\zwom;
		\rfloat = \rmodule*1.2;
		\xmount = \rmodule * 1.3;
		\zscale = \rmodule*4.29;
		\xline = \xmount + .075;
		\ymodule0=0; \ymodule1=5;
		\yline0=\ymodule0-1.5; \yline1=\ymodule1+1.5; 
		\zmodule = 0.;\zmoduleg = 0; \zs = 0.; \cutpos=0;
		} 
	
    \coordinate  (Origin) at (0, 0) ; 
    \coordinate  (labelx) at (-\xmount-4,   0) ;
    \draw[thick, mygraydark] (\xline,0) 
                                            -- (\xline, 13*\zscale);

    \foreach \zmodule\name [count=\counter from 2] in {120/WOM}
    {
    		\pgfmathsetmacro\zs{\counter * \zscale};
    		\pgfmathsetmacro\zmoduleg{int(\zmodule-2660)}
    		
        	\draw [ultra thick, mycolor0, fill=mycolor0b] (\rmodule, \zs+\zwomshift) arc [start angle=0, end angle=180, radius=\rwom] -- +(0,-\zwom) arc [start angle=180, end angle=360, radius=\rwom] -- +(0,\zwom);
    		\draw[very thick, mygraydark] (\xmount,\zs+\zwomshift+\rwom) -- +(0,-2*\rwom-\zwom);
    	 	\draw[ultra thick, mygraydark] (\xmount,\zs+\zwomshift-\zwom*.1) -- +(-3*\rwom,0);
    	 	\draw[ultra thick, mygraydark] (\xmount,\zs+\zwomshift-\zwom*.9) -- +(-3*\rwom,0);
    		\draw (labelx |- 0,\zs) node[align=left,  right]{\name\\ \SI{\zmoduleg}{\meter} $|$ \SI{\zmodule}{\meter}};
    		
    		\pgfmathsetmacro\cutpos{\zs-.45*\zscale};
    }
    
    \foreach \zmodule\name [count=\counter from 3] in {144/PMT-Spectrometer 1,  168/LiDAR 1,  240/Standard Module 1,  264/Mini-Spectrometer,  288/Muon Tracker, 312/Standard Module 2,  384/Standard Module 3,  408/PMT-Spectrometer 2,  432/LiDAR 2}
    {
    		\pgfmathsetmacro\zs{\counter * \zscale};
    		\pgfmathsetmacro\zmoduleg{int(\zmodule-2660)}
    		
        	\draw [ultra thick, mycolor0, fill=mycolor0b] (\rmodule, \zs) arc [start angle=0, end angle=360, radius=\rmodule];
    		\draw[very thick, mygraydark] (\xmount,\zs-\rmodule*1.5) -- +(0,\rmodule*2.7);
    	 	\draw[very thick, mygraydark] (\xmount,\zs+\rmodule*.2) -- +(-2.*\xmount, 0);
    		\draw[very thick, mygraydark] (\xmount,\zs-\rmodule*.2) -- +(-2.*\xmount, 0);
    		
    		\draw (labelx |- 0,\zs) node[align=left,  right]{\name\\ \SI{\zmoduleg}{\meter} $|$ \SI{\zmodule}{\meter}};
    		
    		\pgfmathsetmacro\cutpos{\zs-.45*\zscale};
    }
    \pgfmathsetmacro\zs{12 * \zscale};
    \pgfmathsetmacro\cutpos{\zs-.45*\zscale};
   
     \draw [ultra thick, mycolor1, fill=mycolor1] (\xline+\rfloat, \zs+\rfloat*3.2) arc [start angle=0, end angle=360, radius=\rfloat];
     \draw [ultra thick, mycolor1,fill=mycolor1] (\xline+\rfloat, \zs+\rfloat) arc [start angle=0, end angle=360, radius=\rfloat];
      \draw (labelx |- 0,\zs) node[align=left, above right]{2 Floats\\ \SI{-2216}{\meter} $|$ \SI{444}{\meter} \\ (surface $|$ seafloor)};
      
       \draw[draw=none, rounded corners=1, fill=mygray] (\xline-\rmodule*.5,.5*\zscale-\rmodule) rectangle ++(\rmodule,\rmodule*3);
      \draw (labelx |- 0,.7*\zscale-\rmodule) node[align=left, above right]{Junction box\\ \SI{-2658}{\meter} $|$ \SI{2}{\meter}};
      
      \draw[draw=none, rounded corners=1, fill=mybrown] (\xline-\rmodule*1.5,-\rmodule) rectangle ++(\rmodule*3.,\rmodule*1.3*.5);
      \draw[draw=none, rounded corners=1, fill=mybrown] (\xline-\rmodule*1.5,-\rmodule+\rmodule*1.3*.5) rectangle ++(\rmodule*3.,\rmodule*1.3*.5);
      \draw (labelx |- 0,-\rmodule) node[align=left, above right]{Anchor\\ \SI{-2660}{\meter} $|$ \SI{0}{\meter}};
 
\end{tikzpicture}

%% file: strawb/figure_internal_electronics.tex
\begin{tikzpicture}[every label/.append style={text=black, font=\normalsize},
                                every node/.append style={font=\normalsize}]
  \node[anchor=south west,inner sep=0] (image) at (0,0) {\includegraphics[width=\textwidth,trim={5cm 0 7cm 0},clip]{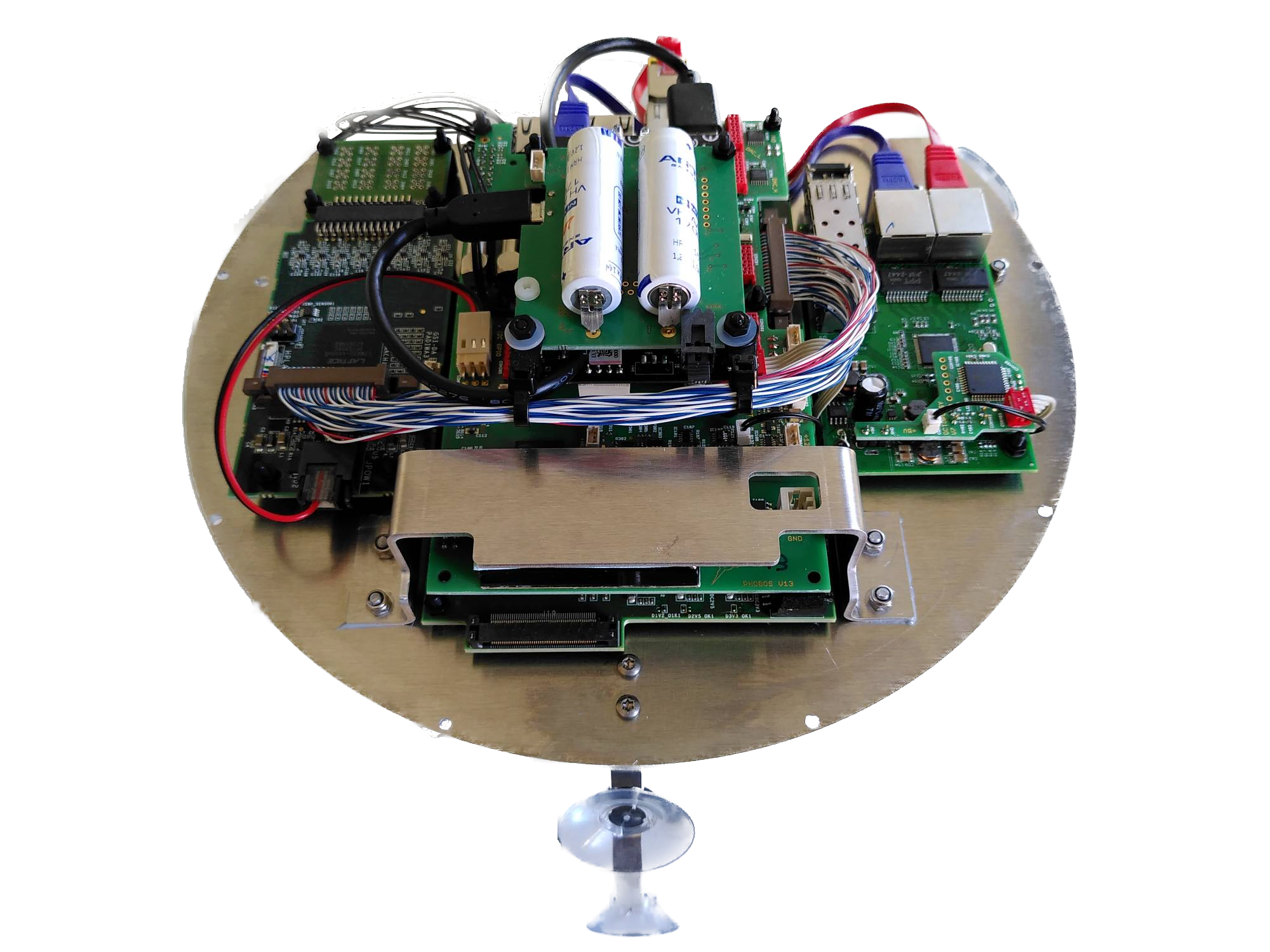}};
  \begin{scope}[x={(image.south east)}, y={(image.north west)}]
  
    \def\xl{0}
  	\node [anchor=west] (padiwa) at (\xl,.875) {Padiwa};
  	\draw [-latex, thick, myarrowcolor0] (padiwa.south) -- ++(0,-.05) -\ (.17,.76);
  	
  	\node [anchor=north west] (logger) at (.1,1) {Logger};
  	\draw [-latex, thick, myarrowcolor0] (logger.east)  -- ++(.05,0) -\ (.43,.84);
  	
  	\node [anchor=west] (odroid) at (\xl,.25) {Odroid};
  	\draw [-latex, thick, myarrowcolor0] (odroid.east) -- ++(0.1,0) -\ (.4,.6);
  	
  	\node [anchor=north west] (power) at (\xl,.2) {TRB3sc};
  	\draw [-latex, thick, myarrowcolor0] (power.east) -- ++(0.1,0) -\ (.4,.3);
  	
  	\node [anchor=north, align=left] (communication) at (.8,.2) {Communication\\Board};
  	\draw [-latex, thick, myarrowcolor0] (communication.north) -| (.8,.5);
  \end{scope}
\end{tikzpicture}

%% file: deployment/deployment.tex
\section{Deployment}
\label{sec:deployment}
As deployment, we describe the procedure of bringing a detector structure from a ship to the sea and placing it on the sea floor at its foreseen location \cite{rules_standards}. For STRAW-b, it also includes the subsequent connection to the NEPTUNE network \cite{onc2013, onc2020}.

Deployments in the deep-sea presents a variety of challenges that stem from the unique, remote, and challenging environment of the ocean depths. The process involves overcoming significant logistical hurdles, as accessing and operating equipment at great depths is inherently complex. The extreme water pressure, low temperatures, and corrosive nature of seawater require the development of specialized materials and robust engineering to ensure the durability of the detectors \cite{iso13628-6, rules_standards}. Additionally, the remote and inaccessible nature of deep-sea locations poses difficulties in deployment and maintenance. Unlike surface installations, where repairs and adjustments can be readily performed, the deep-sea offers limited opportunities for intervention due to its inaccessibility and the high cost associated with underwater operations. Moreover, the light absorption within the water introduces an additional level of complexity, requiring cable connections to provide both power and communication capabilities for the detectors. Overcoming these challenges demands a multidisciplinary approach, combining expertise in marine engineering, materials science, and particle physics \cite{iso13628-6, rules_standards}.

Full-scale neutrino detectors are especially complex as the targeted instrumentation volume is on the scale of one cubic kilometer or larger \cite{km3net2008}. First endeavours of deep-sea neutrino telescope installations where made by the DUMAND project \cite{dumand1992, dumand1995} in the Pacific Ocean. Years after, deep-sea neutrino experimentation predominantly occurred in the Mediterranean Sea, with detectors such as ANTARES \cite{antares2004, antares2011}, NESTOR \cite{nestor2006}, NEMO \cite{nemo2004, nemo2006, nemo2007, nemo2016}, and KM3NeT \cite{km3net2008, km3net2020}. In 2008, ANTARES, NESTOR, and NEMO collaborated to establish and operate the cubic kilometer-scale KM3NeT, pooling their resources to tackle the inherent complexities \cite{km3net2008}. Despite the passage of 16 years, no readily available technology exists for the direct construction of other detectors due to the intricate challenges involved. Consequently, the STRAW-b initiative aims to accumulate valuable experience in preparation for the P-ONE project \cite{pone2023}.

\subsection{Deployment Approach}
Deployments of instrumented mooring lines can be categorized into two primary approaches: top-down and bottom-up \cite{rules_standards}. The bottom-up method involves placing an enclosed structure on the ocean bed and unfolding the mooring line through release mechanisms and its buoyancy. On the other hand, at top-down procedures, the mooring line is lowered from the sea surface to the seabed, typically with a heavy lift line from a deployment vessel. In top-down approaches, the deployment order can be anchor first or buoy first, referring to the part initially deployed from the ship \cite{rules_standards}.

When using a buoy first approach, the mooring line must first be expanded on the sea surface, increasing the deployment time \cite{rules_standards}. Conversely, anchor-first deployments place greater mechanical stress on both the mooring line and the equipment, as the mooring line is responsible for supporting the weight of the anchor throughout the operation. Unfavorable sea conditions can escalate this strain further, necessitating a winch with appropriate load rating on the ship and increasing overall handling complexity. Apart from the deployment method, several other factors must be considered during the process. The sea state, wind speed, and swell are crucial as they impact the positioning of the deployment vessel, as well as the handling of deck work by the crew, and the impact on scientific instrumentation. Therefore, the deployment procedure must account for these factors and adapt to changing weather conditions, even if tailored to a specific sea state. Typically, deployment operations are scheduled during daylight hours with a stable weather window of at least 12 hours \cite{rules_standards}. However, marine operations and equipment handling must be designed to withstand the anticipated loads and sea conditions while ensuring practicality and safety \cite{iso13628-6, rules_standards}.

In the case of \ac{strawb}, a buoy-first deployment was chosen based on a combination of factors, including the available deck space and winches on the ship, as well as the design considerations detailed in \autoref{deployment:sec:design}. Additionally, this approach enables the crew to provide enhanced protection for the delicate instruments by minimizing contact with the ship, as the steel line is under less tension compared to an anchor-first method.

\subsection{Design Considerations}
\label{deployment:sec:design}
The \ac{strawb} project was designed around 14 existing deep-sea cables that were provided from a previous experiment. These hybrid cables are composed of copper wires for power and fiber elements for data transmission, and have been terminated with a dry-mate connector with a length of approximately \SI{20}{\centi\meter}. This design, however, is sub-optimal for the used instrument pressure housing made from glass. The lever arm created by the long connector posed a significant threat to the integrity and durability of the systems. As part of the risk mitigation, a buoy-first top-down deployment procedure was chosen, allowing immediate response by the deck crew during the deployment procedure. The hybrid cables started with a length of \SI{120}{\meter} and increased in steps of \SI{24}{\meter} up to a maximum size of \SI{432}{\meter}. Each of the ten deployed instruments has been connected with its own hybrid cable to the \ac{mjb}, resulting in a star-like network topology. The remaining four connection cables were used for testing. 

\begin{figure}
\includegraphics[width=.662\linewidth]{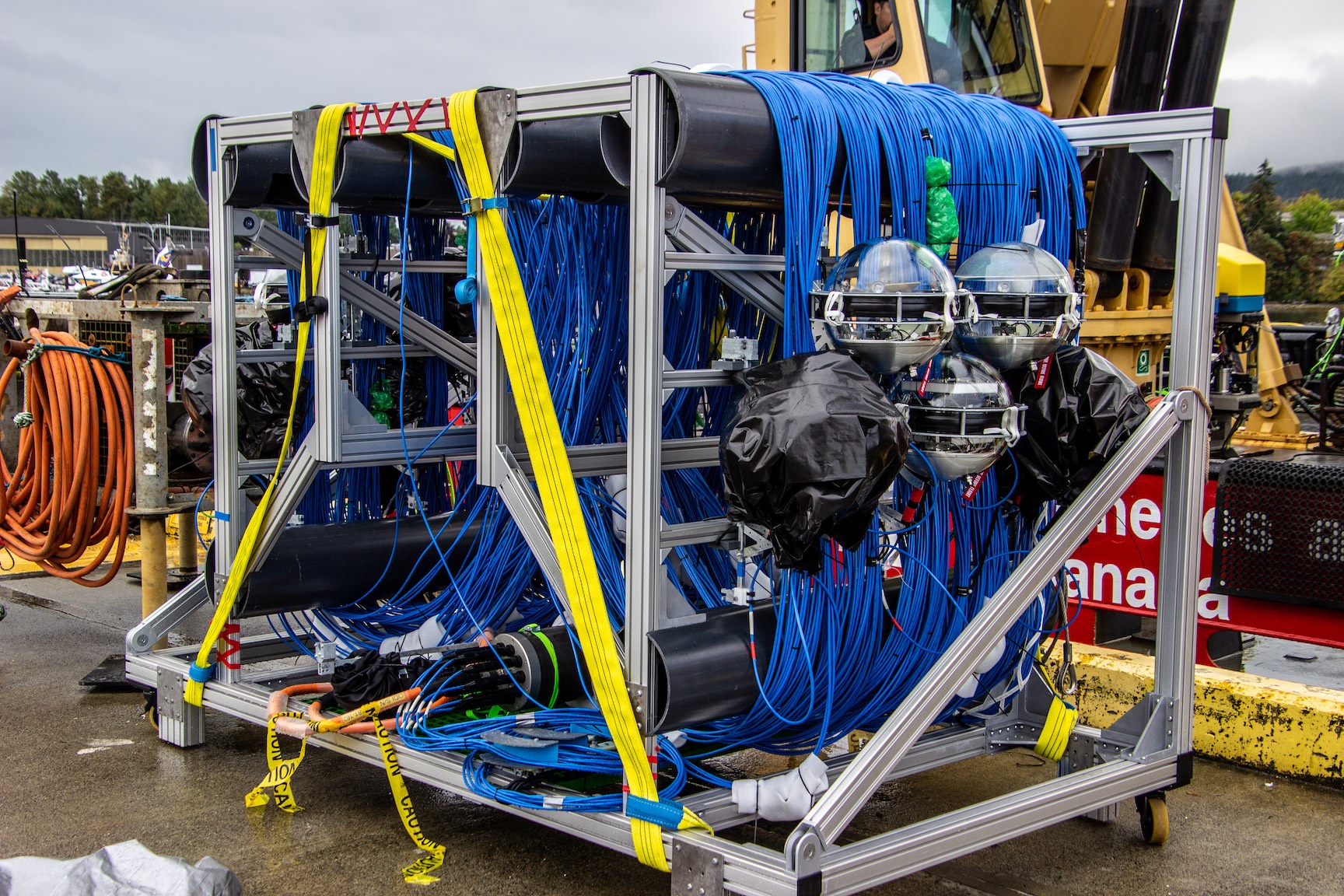}\hfill
\includegraphics[width=.333\linewidth]{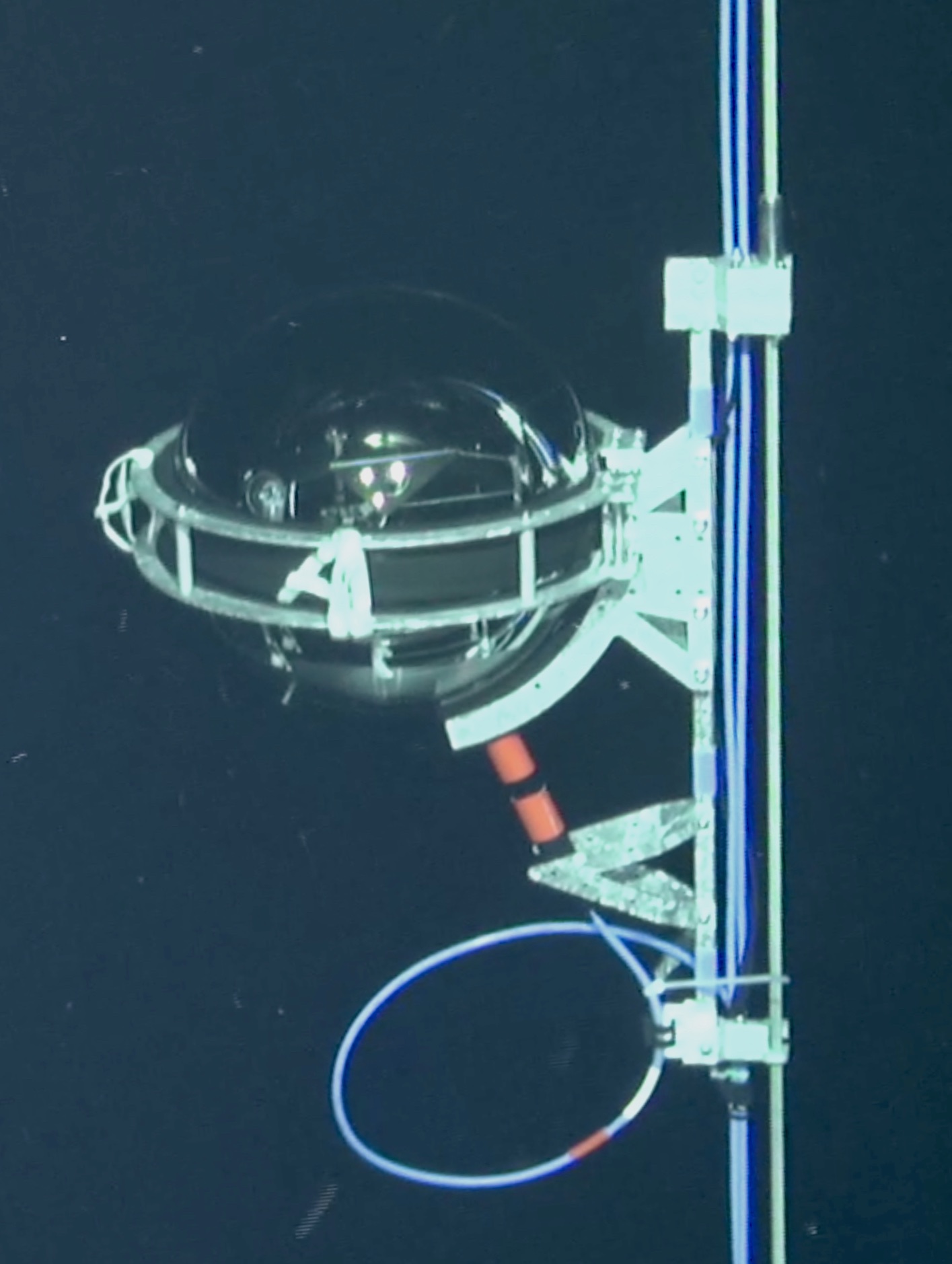}
\caption{Images of the instrument tray with affixed modules (left) and a deployed module (right), captured during the initial inspection dive of a \ac{rov} following the successful deployment.}
\label{deployment:fig:tray}
\end{figure}

The modules, connected to their respective hybrid cable, are stored on an instrument tray (\autoref{deployment:fig:tray}), while the steel line is separately stored on an electric winch. The two components are joined together during deployment on the back deck before going overboard. The mating between the two occurs using a specially designed 'click-in' mechanism with ferrules on the steel line.

The \ac{strawb} module mounting must be a complement to these ferrules, while providing also a secure fixation of the glass sphere without obstructing the field of view of the upper hemisphere. Additionally, the external mounting must be lightweight, as the modules need to be handled during the offshore operation. In addition, the aforementioned difficulties caused by the length of the inherited dry-mate connector had to be taken into account. Several design iterations and module mounting prototypes were required to meet all these requirements. In the end, the design shown in \autoref{deployment:fig:tray} was chosen. Most of the components are laser and water-cut steel parts that can be manufactured in a time and cost efficient manner. The mounting principle of the sphere is based on two steel rings located 30 mm above and below the equator. This ensures that the sealing and adhesive tapes are not affected by mechanical stress. The rings were designed with a small gap between the glass and the steel, which was eventually filled by an ethylene-propylene-diene monomer (EPDM) rubber edge protector. EPDM is a seawater resistant synthetic rubber \cite{ruber}. The external mounting relies on the squeezing of the rubber band to exert sufficient pressure on the sphere to hold it firmly in place. The EPDM rubber is strong enough to protect the glass from damage \cite{spannfellner2020}.

\begin{figure}
\includegraphics[height=.5\linewidth]{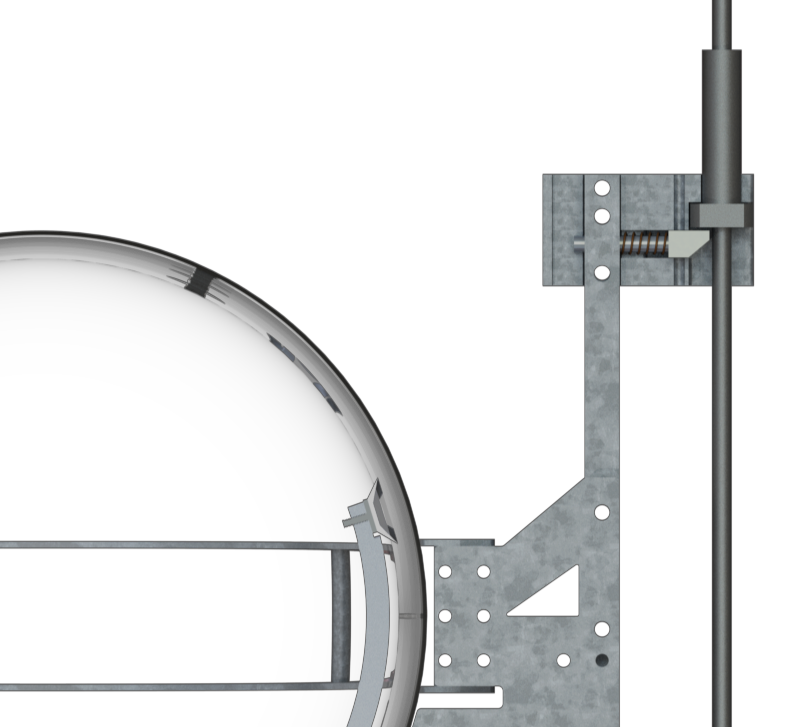}
\includegraphics[height=.45\linewidth]{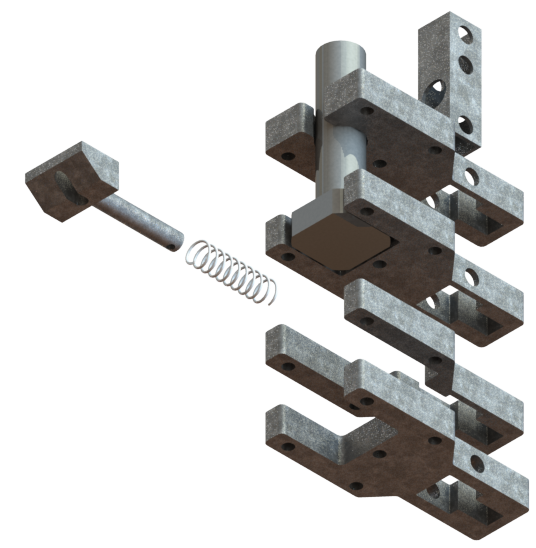}\hfill
\caption{Exploded view of the click-in mechanism (left) and cutaway view when connected to the ferrule on the steel line.}
\label{deployment:fig:click_in}
\end{figure}

The other important aspect of the mounting is the click-in mechanism for the cable ferrules. This is achieved by using a stack of four 12~mm thick steel plates with different recesses to follow the shape of the ferrule. The basic design is shown in \autoref{deployment:fig:click_in}. By using a compression spring and a steel slider, similar to a door lock, the ferrule can be clicked into the mechanism. The sub-components of the external mounting are connected by a square profile. One of the holes in the profile is used as a stop for the compression spring. The square profile has a total length of 650~mm to accommodate the dimensions of the glass sphere and the attached connector. However, as the sphere is offset from the steel cable, it is necessary to prevent the module from tilting. Two small hooks are therefore located at the bottom of the bracket. By rotating the module orthogonally to the bearing steel cable, they can be hooked into and the module can be rotated again to lock the ferrule. As this has to be done on the deck of the support vessel, the mounting had to be lightweight. The total weight of the mount was reduced to around 5~kg by using recesses and struts rather than full metal parts. The strain on the hybrid cable is reduced by a double system: the connector attached to the sphere has static strain relief by means of a cable clamp, while the remaining bundle of cables has a degree of freedom by means of a bungee cord. The maximum elongation of the bungee cord prevents damage to the hybrid cable and connector in the event of increased stress, even when submerged. Protective plates under the ball further protect the connector from mechanical shock. The plates are designed in a triangular shape to allow the bungee cord to slip and prevent knots when the cord is not under tension \cite{spannfellner2020}.

\subsection{Qualification of the Module Mechanics}

To address the challenges associated with the STRAW-b modules, risk mitigation measures have been implemented through a series of tests. These tests specifically focused on evaluating the modules' resistance to environmental conditions in the deep-sea and the stresses encountered during deployment. They were conducted in alignment with standards and recommendations applicable to submarine infrastructure to ensure compliance and reliability \cite{iso13628-6, rules_standards}.

The drag caused by sea currents and the speed of descent during deployment can cause vibration in the instruments, with the potential to damage both electronics and mechanics  \cite{holzapfel2023, iso13628-6, rules_standards}. Vibrations and shocks can also occur during transport and general handling of the equipment. For this reason, vibration and shock tests were carried out on a standard module and a prototype of the PMT spectrometer module in accordance with the ISO~13628-6 standard \cite{iso13628-6}. The test includes a sinusoidal vibration for each spatial axis with an amplitude of 2~mm for the frequency range 5-25~Hz and \SI{5}{\gforce} (\SI{1}{\gforce}=\SI{9.81}{\meter\per\square\second}) between 25-150~Hz. Shocks with an acceleration of \SI{10}{\gforce} are applied eight times per direction \cite{iso13628-6}. Resonance searches within the tested frequency range of 5-150~Hz were also performed before and after the actual tests. Both modules were tested with the external mount, but the square rod and click-in mechanism had to be replaced with an adapter for the test table. Both units passed the test, as did the included readout electronics. The only observed problem was the loosening of four screws at the interface of the readout electronics to the internal mounting plate in one module. This issue was solved in the final design with longer screws and additional use of screwlock \cite{spannfellner2020}.

The pressure in the deep-sea has to be handled by the submersible instruments. For the Cascadia Basin, a pressure of roughly 270~bar is expected at a depth of 2660 m. The glass spheres used are already pressure rated, but their behaviour can be altered by adding specific holes, e.g. for bulkhead connectors. For this reason, a separate pressure test was carried out for a module assembly. The enclosure manufacturer Nautilus provides a special pressure chamber \cite{nautilus} for this purpose. Here the unit is repeatedly cycled to a maximum pressure of 375~bar, before the pressure is held for several minutes. The same module assembly was used for the pressure test as for the vibration and shock tests. The module passed the test successfully, although small detachments of the optical gel from the aluminium shell were observed. However, these quickly disappeared as the vacuum in the module was reduced and are not considered to be relevant to the behaviour of the instrument \cite{spannfellner2020}.

Between the end of December 2019 and the end of January 2020, a STRAW-b module was subjected to a long-term test. It was immersed in a test pool while pressure, temperature and humidity were monitored. The pool was filled with artificial seawater to investigate possible effects on the sealing of the module and connector. Throughout the test period, load resistors were used to simulate a working module. The internal pressure remain fairly stable at around 275~mbar. A steady decrease in humidity was also observed, most likely due to condensation as the module was not purged with nitrogen. Finally, no leakage was detected \cite{spannfellner2020}.

\subsection{Deployment with Data Loggers}
\label{deployment:sec:logger}

\begin{figure}
\centering
\input{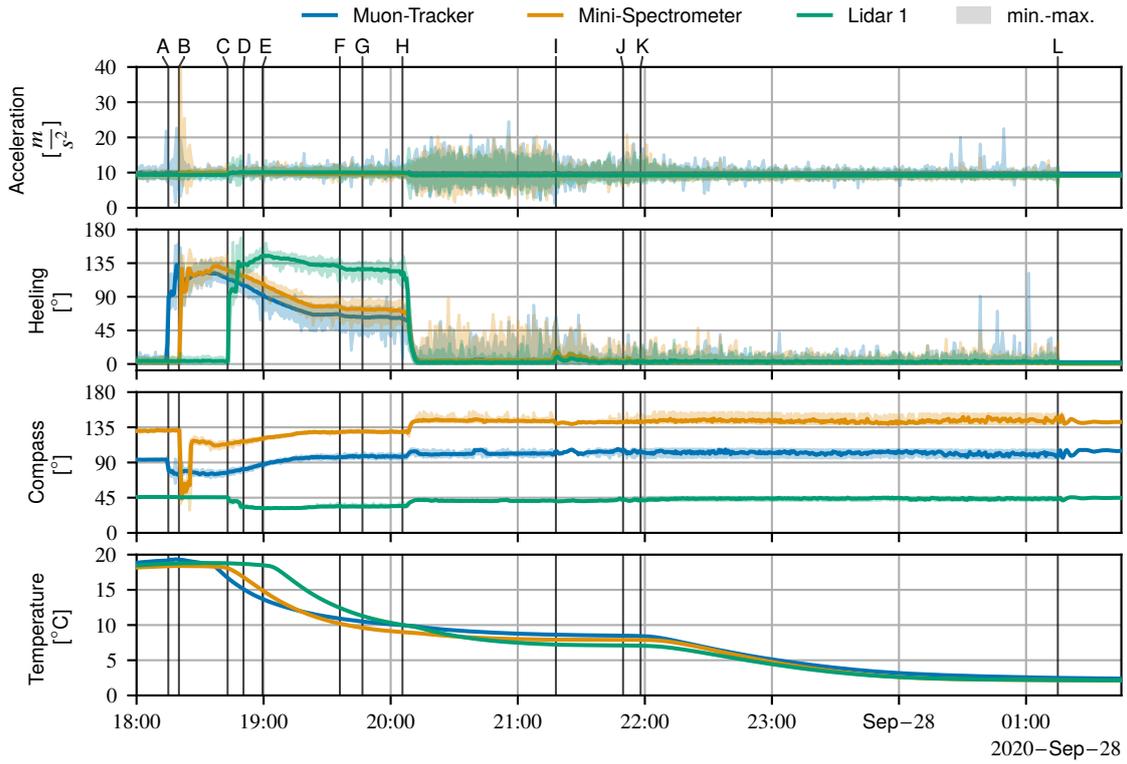}
\caption[Deployment logger data.]{Data obtained from the deployment logger of three modules. The three modules get connected to the string in the order Muon-Tracker, Mini-Spectrometer, and LiDAR~1, indicated as (A-C), respectively. The modules float horizontally behind the ship until the anchor is deployed (H). Afterward, the string starts floating vertically until the ship connects its lift line (I) and begins lowering the string (K). Ultimately, the string is released (L) and descends the remaining $\sim$\SI{5}{\meter} to the seafloor. A complete summary of the marked events (A-L) is provided in \autoref{deployment:sec:logger}.}
\label{deployment:fig:logger}
\end{figure}

In order to quantify the stress on the equipment during deployment, a battery-powered logger was installed in each module, except the \ac{wom}, to assess this process in detail. This logger recorded various parameters including module acceleration, compass heading and internal conditions such as temperature, pressure and humidity. It began recording data after a magnet attached to the instrument casing was removed just before the start of the deployment. The data collected can serve as a valuable reference for future instrument design and testing. \autoref{deployment:fig:logger} shows the data from three loggers. The figure includes markers (A-L) corresponding to specific events covered in the following description.

The deployment procedure commences by attaching three floats to the upper section of the steel line and extending the mooring horizontally on the sea's surface behind the ship, utilizing both currents and a \ac{rhib}. As the steel line is being unspooled, the modules, along with the hybrid cables, are connected to the line on the ship's rear deck. The first connected module is the LiDAR~2, while \autoref{deployment:fig:logger} (A-E) symbolizes the connection of the MuonTracker, Mini-Spectrometer, LiDAR~1, and PMT-Spectrometer~1, respectively. After a module is connected, it is carefully guided into the water to avoid any contact with the boat.
The \ac{mjb} is then affixed to the line (F), followed by disconnecting the line from the winch (G) to mount the anchor. Upon releasing the anchor (H) from the ship's deck, the string aligns vertically and floats freely in the sea. Initially, it was planned to submerge the anchor with a lift line and an acoustic release, but the harsh weather conditions did not allow this step. Afterward, the ship established a connection to the top of the string (I) to remove the extra float and attach the heavy lift line with the acoustic release. The string is gradually lowered, causing the upper float to submerge in the water (J). At this stage, the process briefly halts to attach a beacon for positioning to the lift line. Subsequently, the string commences its descent (K), maintaining a controlled speed of 0.3~m/s. Lastly, the acoustic release is activated, detaching the mooring from the lift line (L), thereby enabling the mooring to descend freely the remaining distance of less than 5 meters to reach the seafloor. During a dedicated \ac{rov} operation the string is inspected and a connection between the pathfinder and the NEPTUNE infrastructure is established, concluding the deployment process.

The logger data clearly reveals the aforementioned distinct phases of the deployment process. The provided numerical values in the following pertain to all loggers, whereas \autoref{deployment:fig:logger} displays a subset of three loggers chosen for clarity. The most noteworthy instances of acceleration are observed when handling the module on the deck (A-E), reaching accelerations of up to \SI{43.4}{\meter\per\square\second}. During the phase where the string is vertically floating on the sea surface, there is an extended period of intense acceleration, reaching levels of up to \SI{30.9}{\meter\per\square\second}. Equally strong is the acceleration seen during the anchor release (H), characterized by a singular peak reaching up to \SI{30.3}{\meter\per\square\second}. The process of lowering the string exhibits a moderate pattern, punctuated by a few brief events of up to \SI{22.5}{\meter\per\square\second} towards the end of the descent. The final drop experiences peaks at \SI{12.8}{\meter\per\square\second}, likely due to the soft seafloor. For comparison, KM3NeT adopts a bottom-up approach, with the reported maximum acceleration during the unfurling process standing at \SI{2.5}{\gforce} (\SI{24.5}{\meter\per\square\second}) \cite{km3net2020}. However, the acceleration experienced during deck operations and the process of lowering the infrastructure to the sea floor is not reported.

The heeling, referring to the angle between the vertical axis and the module's orientation, indicates that it takes approximately \SI{3}{\minute} for the line to align vertically after the anchor is released (H). The compass data shows continuous oscillations with an amplitude of up to \SI{15}{\degree} during the descent (K-L). On the other hand, the temperature data shows a slow decrease to \SI{3}{\celsius} at the seafloor, exhibiting two exponential decays as the module descends segmented into two phases (H, K-L).

%% file: module_operation.tex
\section{Operation of the Instruments in \ach{strawb}}
\label{moperation:sec:strawb_measurement}

\begin{table}
\caption{Summary of \ac{strawb} measurements including readout rate, sensor type, and hosting module.}
\label{moperation:tab:daq}
\begin{tabularx}{\textwidth} {lXll} \toprule
Sensor   & Modules & Rate & Measurement \\ \midrule
3-Axis Accelerometer & All Modules & \SI{0.1}{\hertz} & continuously \\
3-Axis Magnetometer & All Modules & \SI{0.1}{\hertz} & continuously \\
PTH\footnote{PTH: Pressure-Temperature-Humidity sensor.\label{ftn:daq0}} & All Modules & \SI{0.1}{\hertz} & continuously\\
Powermeter\footnote{The Powermeter measures each channel's current and voltage separately.\label{ftn:daq1}} (6x) & All Modules &  \SI{0.1}{\hertz} & continuously \\
Temperature (3x) & All Modules & \SI{0.1}{\hertz} & continuously \\
Camera & Mini- \& PMT-Spec.  & $\sim$1/\SI{90}{\hertz} &  continuously \\
Mini-Spectrometer & Mini- \& PMT-Spec.  & $\sim$1/\SI{90}{\hertz} &  continuously \\
Mini-Spectrometer\footnote{Used for laser calibration before and after each LiDAR measurement.\label{ftn:daq2}} & LiDAR &  2 times\footref{ftn:daq2} &  hourly \\
16 \acs{pmt}s (\ac{tot} counts) & PMT-Spec.  & \SI{1}{\kilo\hertz} &  continuously \\
16 SiPMs (\ac{tot} events) & Muon-Tracker & $>$\SI{1}{\giga\hertz}\footnote{Only the events are stored. The time resolution of the \ac{tdc} is below \SI{1}{\nano\second}. \label{ftn:daq3}} &  continuously \\
LiDAR (\ac{tot} events) & LiDAR & $>$\SI{1}{\giga\hertz}\footref{ftn:daq3} &  hourly \\

\bottomrule
\end{tabularx}
\end{table}

The \acf{daq} system of \ac{strawb} records the data either continuously or on specific schedules. All modules contain ambient sensors to monitor the conditions inside the pressure housing, while some modules host additional sensors, as outlined in \autoref{strawb:sec:strawb}. This section presents a summary of the various measurements, while \autoref{moperation:tab:daq} provides a general overview, including readout rate, sensor types, and corresponding hosting modules. All recorded data from the various measurements are made accessible to the public through a database known as \textit{Oceans 2.0}, or its more recent version, \textit{Oceans 3.0}, as described in \autoref{daq:sec:oceans}. Inside this database, files are methodically organized using data-product codes that align with the specific measurements they represent. Subsequent sections will provide specific information regarding these data-product codes.

\subsection{Standardized Module}
The standardized module, which each module builds on, is equipped with a 3-axis accelerometer and a 3-axis magnetometer that provide information on movement and orientation. Additionally, the module includes sensors for monitoring pressure, humidity, temperature, and power consumption within the pressure housing. The \ac{daq} also keeps track of various software and system settings. Sensor data is collected every \SI{10}{\second}, while software and system settings are only saved when changes are made to reduce unnecessary data. The collected data is labeled with the data-product-code \textit{SMRD}.

\subsection{Cameras, Mini-Spectrometer, and LEDs}
The Mini-Spectrometer and the two PMT-Spectrometer modules contain a camera and a mini-spectrometer from Hamamatsu \cite{minispec2023}. These sensors synchronously operate as they share the same \ac{spi}-bus for communicating. The camera has a readout limitation, resulting in one image and spectrum being recorded every \SI{99.5}{\second}. Moreover, the PMT-Spectrometer 2 captures a set of images with enabled LEDs in the neighboring module (LiDAR 2) once per hour to track its position. The measurement with enabled LEDs ceased after LiDAR 2 became unresponsive in late 2021. The LED activation and configuration information is logged by the \ac{daq} and stored in a file with data-product-code \textit{SMRD}. Additionally, the camera and Mini-Spectrometer each have their own files, which are labeled with data-product-codes \textit{MSSCD} and \textit{MSSD}, respectively.

\subsection{PMT-Spectrometer}
The PMT-Spectrometer continuously records the PMT signals in the low precision mode of the \ac{trb} \cite{trb}, which means counting \ac{tot} events rather than storing the time and duration of each event separately as described more precisely in \autoref{daq:sec:fast_signal}. The \ac{tot} counts are stored with a frequency of \SI{1}{\kilo\hertz}. Once a day (around midnight UTC), a threshold scan has been conducted for the \ac{tot}, interrupting the continuous measurement for less than five minutes. The \ac{trb} is rebooted during the threshold scan. The \ac{daq} system continuously monitors communication, and if an issue arises, it triggers a reboot of the \ac{trb} and restarts the measurement. The PMT-DAQ files, with the data-product-code \textit{PMTSD}, rotate every hour to ensure file sizes remain below 100 MB. The utilization of PMT-Spectrometer 2 is precluded for this measurement due to a malfunctioning FPGA, which hinders the retrieval of data from the \acp{pmt}.

\subsection{Muon-Tracker}
The Muon-Tracker utilizes the high precision mode of the \ac{trb} \cite{trb}, as described in \autoref{daq:sec:fast_signal}, to track photon detection in the scintillators with the SiPM. In this mode, all \ac{tot} events, i.e., events with timestamps and durations over the threshold, are stored. Due to the fine-tuned threshold and subsequently low background of the SiPM arrays on the scintillators, the files rotate only once a day, remaining below 100 MB, with the data-product-code \textit{MTSD}. As with the PMT-Spectrometer, a threshold scan is performed once a day, and the \ac{daq} system continuously monitors the \ac{trb}, triggering a restart of the measurement if needed.

\subsection{LiDAR}
The LiDARs perform various measurements, including a daily scan of the surrounding volume with the gimbal rotating over 2$\pi$. This scan is repeated without the enabled laser to gather information on in-situ afterpulsing characteristics. A daily laser scan is also performed, where the gimbal moves in a fixed direction, and the laser changes its alignment relative to the optical axis of the detection optics. When no other measurements are ongoing, two LiDAR measurements with the enabled laser and one without the enabled laser are taken per hour. During those measurements, events are recorded for \SI{60}{\second}, and the gimbal points in the same direction as during the laser scan. The Laser trigger signal and the PMT's photo counting base signal are both digitized in high precision mode by the \ac{trb}, ensuring that precise time information is recorded for every Laser pulse and incoming photon. For more information about the \ac{trb}, please refer to the details provided in \autoref{daq:sec:fast_signal}. Each measurement is saved in one \ac{hdf5} file identified by the data-product-code \textit{LIDARSD}. In addition, the raw HLD files (can be multiple for a single measurement) from the \ac{trb} are stored and synchronized with the \ac{onc} database.

%% file: daq/daq.tex
\section{Distributed Control System and  Data Acquisition of \ach{strawb}}
\label{daq:sec:daq_file_structure}

The \ac{mctl} operates all \ac{strawb} modules, except the \ac{wom}, and runs on each of them as an autonomous controller implemented as a \acf{dcs} \cite{dcs1995}. Its primary goal is to ensure all sensor's safe and streamlined operation, both in manual and automated modes. To this end, the \ac{mctl} includes various functionalities such as safety control, logging, sensor operation and monitoring, bus management, inter-module communication, \ac{daq}, and measurement execution and scheduling. 

\subsection{Data Acquisition System}

\begin{figure}
\input{daq_file_structure}
\caption[Generalized HDF5 file structure of STRAW-b's DAQ]{Generalized HDF5 file structure of \acs{strawb}'s \acs{daq}. It exemplarily shows three data groups (A, B, C), each with two datasets and the time dataset.}
\label{daq:fig:daq_file_structure}
\end{figure}
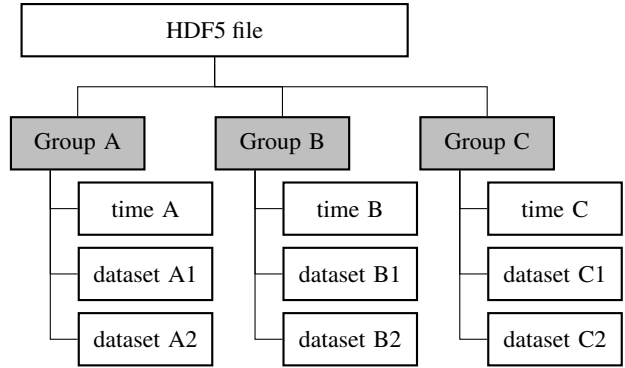

The \ac{daq} collects all sensor values and stores the data. To seamlessly integrate with the deep-sea infrastructure, a specialized software python package named SDAQ (Simple Data Acquisition) has been created and is accessible to the public \cite{sdaq}. This package is optimized for time series data stored in \ac{hdf5} files using lossless data compression and checksums natively supported by \ac{hdf5} \cite{hdf5}. The standardized file structure employed by the \ac{daq} is illustrated in \autoref{daq:fig:daq_file_structure}. Its purpose is to systematically arrange time series data, which are essentially sets of measurements taken over a period. Within the file, each \ac{hdf5} group is dedicated to represent a distinct time series. The primary dimension across all datasets within a group is associated with time, and each \ac{hdf5} group contains a dedicated \textit{time} dataset to store the timestamps of the respective time series. This organization ensures uniform sizes for all datasets within a group. For extended or continuous measurements, the \ac{daq} system initializes a file rollover after fixed periods, such as hourly or daily. In addition, metadata are stored in the attributes of the \ac{hdf5} file, group, and datasets. For example, in the context of continuous measurements, each file is assigned a unique ID, and this identifier is stored as a file attribute along with the ID of the preceding file. This approach allows for a simplified retrieval of continuous measurements at a later point in time.

\subsection{Integration into Oceans 2.0/3.0}
\label{daq:sec:oceans}
After recording the data on the modules, the data are downloaded and integrated into \textit{Oceans 2.0}. \textit{Oceans 2.0}, or its newer iteration 3.0, serves as a publicly accessible \ac{db} for all experiments connected to the \ac{onc} infrastructure. \ac{onc} manages this \ac{db}, and it can be accessed through the webpage at \url{https://data.oceannetworks.ca/home} or via \acp{api} \cite{oncOceans2020}. Within this \ac{db}, files are systematically organized and labeled with data-product codes that correspond to their specific measurements. The data-product codes associated with \ac{strawb} are detailed in \autoref{moperation:sec:strawb_measurement}.

\ac{mctl} is equipped with an interface to \ac{onc}'s shutdown procedure in case the modules need to be powered off. In such a scenario, \ac{mctl} stops all measurements, stores the data, and puts the module in a safe state concerning software and hardware. The final shutdown command to the operating system is triggered by \ac{onc}'s shutdown procedure.

\subsection{Communication Management}
Some electronic components share a communication bus where only one member can send data at a time, such as \aci{spi} or \aci{i2c}. The bus-management function keeps track of all communications and distributes the available resources to prevent interference. The communication between modules depends on the utilization of the network protocol \ac{ssh} and manages measurements in cases where multiple modules are engaged. This communication holds particular significance in safeguarding sensitive PMTs from potential overexposure by adjacent LEDs. Given that the PMTs and LEDs may be situated in separate modules, the module communication system ensures that it only activates the PMTs' \ac{hv} while deactivating the LEDs, and vice versa.

\subsection{Safety Control and Monitoring System}

The safety control prevents any actions that potentially damage the sensors or cause data loss. For example, if an LED is activated in the same or neighboring module, the safety control will block the powering of \acp{pmt} to avoid exposing them to intense light. The logging function generates a detailed record of system events, warnings, and errors, which is essential for effective debugging, especially in case of rare issues. The sensor operation manages low-level functions required to operate individual sensors, while the sensor monitoring ensures their proper functioning and triggers corrective action in case of any malfunctions. 

The monitoring system of each module is integrated into a real-time monitoring system for the entire detector, which uses the open-source tools Telegraph, InfluxDB, and Grafana to collect, store, and visualize sensor data in real-time \cite{timeseries, tig_stack}. The combination of the tools is also called the \textit{TIG Stack} \cite{tig_stack}. Telegraph is used to gather data from various sources, including sensors and services, which are then sent to InfluxDB and stored as time series data.

\begin{figure}
\caption[The real-time monitoring system based displaying the last connection to the LiDAR 2 module.]{A screenshot of the real-time monitoring system based on the \textit{TIG Stack} (Telegraf, InfluxDB, Grafana) \cite{timeseries, tig_stack}, displaying the last connection to the LiDAR 2 module. This system collects sensor data in real-time from various modules and allows for easy visualization and detection of any anomalies through customizable Grafana dashboards.}\label{fig:strawb_tigstack}
\includegraphics[width=.8\textwidth]{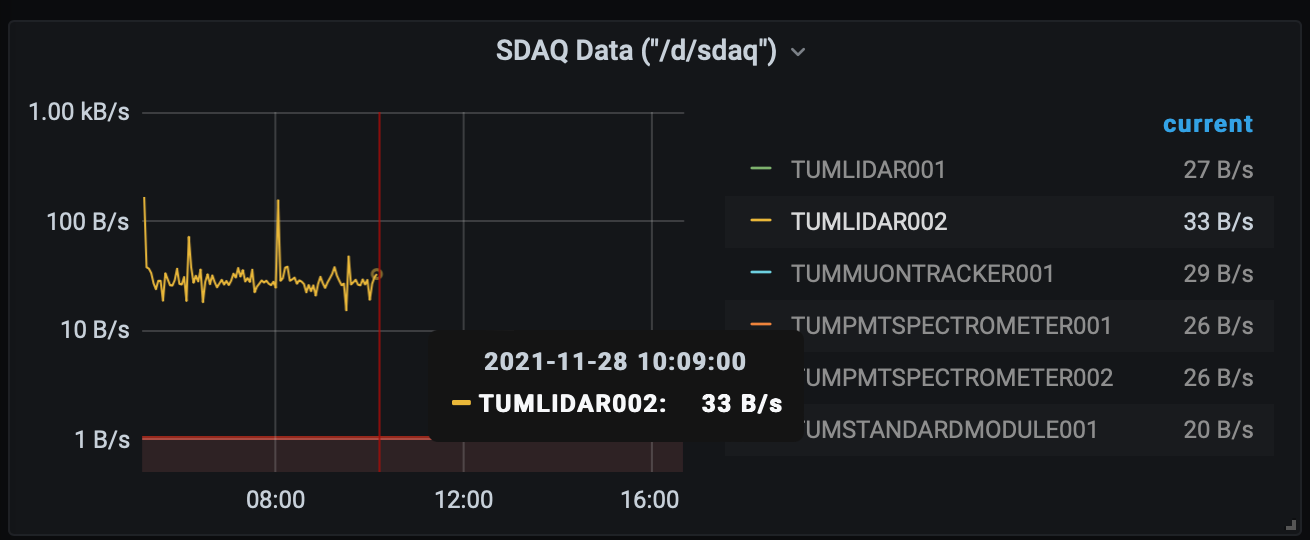}
\end{figure}

Grafana is used to create customizable dashboards to visualize the data in real time, allowing for easy monitoring and detection of any anomalies. Additionally, the system has the ability to send alerts in real-time based on predefined thresholds, ensuring that any issues are addressed promptly \cite{timeseries, tig_stack}. \Autoref{fig:strawb_tigstack} shows the last connection of the LiDAR 2 module to demonstrate the monitoring system's performance. 

%% file: daq/daq_file_structure.tex
\begin{tikzpicture}[
every label/.append style={text=black, font=\footnotesize},
every node/.append style={font=\footnotesize, draw=black,thick,anchor=west, minimum height=1.8em} ,
criteria/.style={text centered, text width=1.5cm, fill=gray!50},
attribute/.style={%
    grow=down, xshift=0cm,
    text centered, text width=1.5cm,
    edge from parent path={(\tikzparentnode.225) |- (\tikzchildnode.west)}},
first/.style    ={level distance=5ex},
second/.style   ={level distance=10ex},
third/.style    ={level distance=15ex},
fourth/.style   ={level distance=20ex},
fifth/.style    ={level distance=25ex},
level 1/.style={sibling distance=7em}]
    \node[anchor=south, minimum width=2in]{HDF5 file}
    [edge from parent fork down]

    child{node (crit1) [criteria] {Group A}
        child[attribute,first]  {node {time A}}
        child[attribute,second] {node {dataset A1}}
        child[attribute,third]  {node {dataset A2}}
        }
    child{node [criteria] {Group B}
        child[attribute,first]  {node {time B}}
        child[attribute,second] {node {dataset B1}}
        child[attribute,third]  {node {dataset B2}}
        }
    child{node [criteria] {Group C}
        child[attribute,first]  {node {time C}}
        child[attribute,second] {node {dataset C1}}
        child[attribute,third]  {node {dataset C2}}
        }
        ;
\end{tikzpicture}

%% file: software_package.tex
\section{A Software Package for the Data Analysis for \ach{strawb}}
\label{strawb_package:sec}

Managing and analyzing large datasets, like those generated by the \ac{strawb} detector, can be a challenging task. To simplify this process, a user-friendly software package called \textit{strawb} has been developed in Python. This package provides tools for synchronizing data from \ac{onc}'s Oceans 2.0/3.0 database to a local database, as well as importing and analyzing the data. Currently, the \textit{strawb} package repository is not public.

The \textit{strawb} package offers several key features, such as the ability to synchronize available data from the \ac{onc} database to a local database. This feature allows users to share their code with others, with the package taking care of synchronizing the required files without downloading all files from the \ac{onc} database. By downloading only the required files, \textit{strawb} minimizes the amount of data transfer and optimizes working with large datasets.
In addition, \textit{strawb} provides basic import functionalities for different file types, making it easy to import and process data from the \ac{strawb} detector. This feature reduces the time and effort required for data management.
It also includes basic and advanced analysis tools that every user can use to streamline their work with the data. These tools include functions for data visualization, quality control checks, and event selection, as well as advanced analysis techniques such as signal processing and machine learning.
Additionally, \textit{strawb} can access other sensors available on the \ac{onc} database, such as current profilers and temperature sensors. It also includes functionalities to access metadata from the \ac{onc} database,  such as position information of the devices.

%% file: summary_outlook.tex
\section{Summary and Outlook}

As stated in this report, the successful deployment of \ac{strawb} was made possible by the careful development and design considerations for the \SI{444}{\meter}-long mooring line, coupled with the deep-sea expertise of \ac{onc}. The battery-powered deployment loggers provided valuable insights into the mechanical stress experienced during the deployment. This information can be utilized for designing and testing equipment in future developments.

Furthermore, the commissioning phase of the \ac{strawb} project concluded in a continuous and automated detector operation for various measurements. Numerous security and monitoring mechanisms have been implemented to ensure robust and uninterrupted performance, guaranteeing high uptimes for all sensors. Notably, the \ac{daq} system has been seamlessly integrated into \ac{onc}'s Oceans 2.0/3.0 database, facilitating the availability of all collected data to the public. This integration enhances accessibility and promotes transparency, allowing researchers and interested individuals to access and utilize the valuable data collected by \ac{strawb}.

The planned and successful recovery of \ac{strawb} during the Nautilus cruise in July 2023 marks the end of the operational life of the mooring line. For the recovery, the specialized winch from the deployment was reused to coil up the wire rope, while the modules could be easily removed with the still operational click-in mechanism. Following the recovery, a thorough analysis of all instruments will be conducted to gather input for designing the detector lines of P-ONE. Preliminary visual examinations indicate that all three module failures where the connections were lost are attributed to leaking connectors on either the \ac{mjb} or module side. This leakage has led to water infiltration between connectors and penetrators, ultimately resulting in salt corrosion. Additionally, biological samples from sedimentation and bio-fouling have been collected, providing supplementary data to complement the optical measurements conducted by pathfinders.

Moving forward, our primary aim involves an extensive analysis of the rich data amassed by \ac{strawb}, focusing on quantities directly pertinent to the functioning of \ac{pone}. An essential aspect under investigation is the impact of sedimentation and bio-fouling on the modules. These factors can significantly degrade the light sensitivity of the modules, which becomes especially crucial over the extended operational lifespan of \ac{pone}, spanning multiple decades. The data sourced from the Muon-Tracker will be harnessed to assess its potential as a calibration tool for \ac{pone}. Moreover, the LiDAR data offers valuable insights into water attenuation and scattering characteristics, providing a complementary perspective to the information gleaned from \ac{straw} measurements. A noteworthy emphasis will be placed on advancing our comprehension of bioluminescence. This pursuit will encompass data analysis from the PMT-Spectrometers, cameras, and Mini-Spectrometers. Additionally, the integration of data from other deployed sensors at the Cascadia Basin, e.g., the current meters operated by \ac{onc}, will contribute to our comprehensive understanding.

%% file: abbreviations.tex

\begingroup
\setstretch{.5}
\begin{acronym}[TDMAWGN]
\acro{adc}[ADC]{analog to digital converter}
\acro{adcp}[ADCP]{acoustic Doppler current profiler}
\acro{ann}[ANN]{artificial neural network}
\acro{agn}[AGN]{Active Galactive Nuclei}
\acro{aov}[AOV]{angle of view}
\acro{api}[API]{application programming interface}
\acro{ccd}[CCD]{charge\--coupled device}
\acro{cfa}[CFA]{color filter array}
\acro{cmos}[CMOS]{complementary metal\--oxide\--semiconductor}
\acro{cpld}[CPLD]{complex programmable logic device}
\acro{cwl}[CWL]{center wavelength}
\acro{daq}[DAQ]{data acquisition}
\acro{db}[DB]{data base}
\acro{dc}[DC]{direct current}
\acro{dcs}[DCS]{distributed control system}
\acro{pdf}[PDF]{probability density function}
\acro{dna}[DNA]{deoxyribonucleic acid}
\acro{dnn}[DNN]{deep neural network}
\acro{efl}[EFL]{effective focal length}
\acro{fft}[fft]{fast fourier transformation}
\acro{fov}[FoV]{field of view}
\acro{fpga}[FPGA]{Field-Programmable Gate Array}
\acro{fwhm}[FWHM]{full width at half maximum}
\acro{hdf5}[HDF5]{Hierarchical Data Format}
\acro{hv}[HV]{high voltage}
\acro{i2c}[I\textsuperscript{2}C]{Inter-Integrated Circuit}
\acro{imo}[IMO]{international maritime community}
\acro{kde}[KDE]{Kernel Density Estimation}
\acro{km3net}[KM3NeT]{Cubic Kilometer Neutrino Telescope}
\acro{led}[LED]{light\--emitting diode}
\acro{lidar}[LiDAR]{light detection and ranging}
\acro{mctl}[MCTL]{master control software}
\acro{mi}[MI]{mutual information}
\acro{mjb}[MJB]{mini junction box}
\acro{ml}[ML]{machine learning}
\acro{msb}[MSB]{marine snow, sedimentation, and bio-fouling}
\acro{ndsc}[NDSC]{network for detection of stratospheric change}
\acro{nn}[NN]{neural network}
\acro{onc}[ONC]{Ocean Networks Canada}
\acro{pmt}[PMT]{photomultiplier tube}
\acro{pocam}[POCAM]{Precision Optical Calibration Module}
\acro{pone}[P-ONE]{Pacific Ocean Neutrino Experiment}
\acro{qe}[QE]{quantum efficiency}
\acro{gvd}[GVD]{Gigaton Volume Detector}
\acro{rhib}[RHIB]{Rigid-Hulled Inflatable Boat}
\acro{rov}[ROV]{remotely operated vehicle}
\acro{sdom}[sDOM]{STRAW Digital Optical Module}
\acro{sipm}[SiPM]{silicon photomultiplier}
\acro{spi}[SPI]{serial peripheral interface}
\acro{ssh}[SSH]{secure shell protocol}
\acro{straw}[STRAW]{Strings for Absorption Length in Water}
\acro{strawb}[STRAW\--b]{Strings for Absorption Length in Water b}
\acro{tdc}[TDC]{time\--to\--digital converter}
\acro{tot}[ToT]{time\--over\--threshold}
\acro{trb}[TRB]{Trigger Readout Board}
\acro{tts}[TTS]{transit time spread}
\acro{utc}[UTC]{coordinated universal time}
\acro{uvc}[UVC]{ultraviolet C}
\acro{uv}[UV]{ultraviolet}
\acro{veoc}[VEOC]{vertical electrical optical cable}
\acro{wom}[WOM]{Wavelength-shifting Optical Module}

\end{acronym}

\endgroup